\RequirePackage{fix-cm}
\documentclass[smallcondensed]{svjour3}     
\smartqed  
\usepackage[title]{appendix}%
\usepackage{colortbl}
\usepackage{tcolorbox, color, soul}
\usepackage[figuresright]{rotating}
\usepackage{ulem}
\usepackage{booktabs}
\usepackage{url}

\newtcolorbox[auto counter]{mybox_old}[1][]{%
    title=Search query (initial), arc=1pt,outer arc=1pt, #1}
\newtcolorbox[auto counter]{mybox_new}[1][]{%
    title=Search query (updated\, final), arc=1pt,outer arc=1pt,
    #1}
\definecolor{Gr}{RGB}{211, 211, 211}	
\definecolor{LightCyan}{rgb}{0.88,1,1}

\begin{document}

\title{Machine Learning Models for the Early Detection of Burnout in Software Engineering: a Systematic Literature Review}




\author{Tien Rahayu Tulili        \and
        Ayushi Rastogi \and
        Andrea Capiluppi
}


\institute{Tien Rahayu Tulili \at
              Bernoulli Institute for Mathematics, Computer Science, and Artificial Intelligence, University of Groningen,
              Nijenborgh 9, 9747AG, Groningen, Groningen, The Netherlands \\
              \email{t.r.tulili@rug.nl, tien.tulili@polnes.ac.id}           
           \and
           Ayushi Rastogi \at
              \email{a.rastogi@rug.nl}
           \and
           Andrea Capiluppi \at
           \email{a.capiluppi@rug.nl}
}

\date{Received: date / Accepted: date}

\maketitle

\begin{abstract}
Burnout is an occupational syndrome that, like many other professions, affects the majority of software engineers. Past research studies showed important trends, including an increasing use of machine learning techniques to allow for an early detection of burnout. 

This paper is a systematic literature review (SLR) of the research papers that proposed machine learning (ML) approaches, and focused on detecting burnout in software developers and IT professionals. Our objective is to review the accuracy and precision of the proposed ML techniques, and to formulate recommendations for future researchers interested to replicate or extend those studies. 

From our SLR we observed that a majority of primary studies focuses on detecting emotions or utilise emotional dimensions to detect or predict the presence of burnout. We also performed a cross-sectional study to detect which ML approach shows a better performance at detecting emotions; and which dataset has more potential and expressivity to capture emotions.

We believe that, by identifying which ML tools and datasets show a better performance at detecting emotions, and indirectly at identifying burnout, our paper can be a valuable asset to progress in this important research direction.

\keywords{burnout \and machine learning models \and software engineering \and systematic literature review}
\end{abstract}

\section{Introduction}
Burnout, a prevalent occupational syndrome among professionals across various fields, manifests itself in diverse forms, with emotional exhaustion, depersonalisation, and a sense of reduced personal accomplishment being widely recognised as prime indicators of this condition~\cite{maslach2006burnout}. This syndrome, like many other professions, affects the majority of software engineers: recent surveys reported that a staggering 82\% of the surveyed software developers suffered from burnout~\cite{SurveyHaystack,SurveyCodeAhoy}.

Individuals experiencing burnout frequently express elevated levels of negative emotions, including feelings of frustration, anger, sadness  and anxiety. The persistent and long-term exposure to stress and work-related pressures can result in the gradual build-up of negative emotions, which in turn contributes to the emergence of burnout. Specifically, the emotional exhaustion aspect of burnout is closely associated with the experience of negative emotions, as individuals may feel depleted, overwhelmed, and emotionally drained~\cite{holmqvist2006burnout,alessandri2018job,schoeps2019effects,liu2021negative}.

On the one hand, in other fields (health, education, and sports), similar risk factors contribute to burnout, particularly the emotional exhaustion dimension, and including environmental (e.g. work climate or time pressure)~\cite{salles2020assessment,kleiner2017oncologist}, and psychological factors (e.g., mental stress, self-efficacy, self-determination)~\cite{Lee2016jpts2016286,dreison2018integrating,makara2019self,ventura2015professional}, regardless of the different measurement indicators of burnout applied. For those studies,  audiotapes or interview's transcripts were mostly used as data source, to analyse communication behaviours and to investigate burnout in health field~\cite{ratanawongsa2008physician,robbins2019provider,akhavan2022going}, human service workers~\cite{miller2007compassionate}, education~\cite{hesham2023special}, and sports~\cite{li2017roles}.

On the other hand, several factors (specific to the Software Engineering field) contribute to increased burnout. The complexity of the work, which demands constant deep focus and abstract thinking, can be mentally draining~\cite{complexitykillingdevs}. The pressure from frequent sprints, tight deadlines, and continuous delivery in Agile and DevOps environments adds relentless stress with minimal downtime~\cite{DevopsBurnout,greatdevopsburnout,kam2023devops}. Additionally, blurred boundaries between work and personal life~\cite{SurveyHaystack} and extended hours in sectors like startups and gaming further contribute to higher burnout levels among software engineers~\cite{peticca2015perils}. 

Other specific aspects, such as industry-specific jargon and specific tools employed during software development, can also play an important role in increasing the risk of burnout. For example, a unique artefact like the Version Control System (VCS), which is useful to manage the changes within a system, may trigger conflicts among developers, especially when they have to merge the changes and integrate them correctly into the system~\cite{bird2012assessing,estler2014awareness}. Resolving the conflicts can be particularly stressful, especially when time is limited~\cite{kasi2013cassandra}. Furthermore, repeated refactoring (especially in large, complex, or poorly designed systems) can be labour-intensive and mentally exhausting~\cite{dig2007refactoring}. In addition, employing an issue tracking system (e.g. JIRA) during software development may escalate task tracking, particularly if the workflow is poorly designed~\cite{bertram2010communication}. This may lead to overwhelming and demoralised feelings.

Past research studies showed important trends, including an increasing use of machine learning techniques to allow for an early detection of burnout. Research in the field of software engineering has investigated the emotional experiences of IT professionals: in particular, studies have aimed to recognise initial signs of burnout by examining emotions shown during software development~\cite{S1_Mantyla2016,S2_Gachechiladze2017}. 


In this paper we focus our literature review on the selection of past research studies that focused on the early identification of burnout. Specifically, we are interested in studies that approached this issue through the detection of emotions, feelings, sentiments, and stress -- factors closely associated with burnout~\cite{rey2016emotional,spiller2021emotion,zapf2002emotion,abellanoza2018burnout} -- using one or more machine learning techniques.
Our motivation is to review which ML approach shows the better results (in terms of performance) when attempting to categorise and identify, at an early stage, the emotions, stress and sentiments that are related to burnout.

This paper contributes to the current literature on burnout in software engineering by presenting:
\begin{itemize}
    \item a set of 64 research studies that proposed ML approaches that can be used as the basis of future investigation in detecting burnout;
    \item a review of the accuracy and precision of the proposed ML techniques; 
    \item a set of recommendations for future researchers interested to replicate or extend those studies.
\end{itemize}

Our paper is articulated as follows: Section~\ref{sec:_related-works} presents the main findings of past SLRs; Section~\ref{sec:_methodology} describes the methodology of our SLR; Section~\ref{sec:_findings_rq1_early} to Section~\ref{sec:_findings_rq5_datasets} outline the SLR findings;
Section~\ref{sec:_discussions_and_implications} presents the discussions and implications.  Section~\ref{sec:_threats} provides the threats to the validity of this SLR and, Section~\ref{sec:_conclusion} concludes.

\section{Related Works}
\label{sec:_related-works}
To the best of our knowledge, an overview of machine learning-based methods in detecting burnout has not been conducted yet. Therefore, in this section we discuss all the secondary studies (e.g. systematic literature studies) that have been focused on sentiment analysis, emotion recognition/detection, and mental disorder detection.

Previous secondary studies have so far focused on sentiment tools for analysing texts retrieved from different high-dynamic data sources, including social media, and software engineering-related media.

For instance, Zucco et al.,~\cite{zucco2020sentiment} conducted a research study that reviewed methods and tools employed for Sentiment Analysis (SA), particularly within established environments like social networks. This study compared 24 tools based on several criteria and variables. Additionally, the analysis and testing of the tools were conducted in the context of usability, flexibility of use, and other specifications appertaining to the type of analysis performed.

Meanwhile, the SLR conducted by Obaidi and Kl{\"u}nder~\cite{obaidi2021development} also discussed the SA tools, but it evaluated them for use in the Software Engineering domain. In this review, the authors focused on the development and the future applicability of the tools, and also  analysed the available Sentiment Analysis methods and tools in various application scenarios.

Lin et al.,~\cite{lin2022opinion} extensively discussed the tools and resources available in the context of opinion mining: the off-the-shelf opinion mining approaches, the available datasets for performance evaluation and tool customisation, and the recommendations considered when adopting or customizing the opinion mining techniques. 

Other SLRs have focused on the data sources used to determine software engineers' moods. The study done by S{\'a}nchez-Gord{\'o}n and Colomo-Palacios~\cite{sanchez2019taking} for example proposed, as alternate data sources for identifying emotions, the self-reported emotions or the readings by biometric sensors.

In addition, the SLR presented by Tawsif~\cite{tawsif2022systematic} focuses on an emotional state `recognition system', that was built using physiological signals obtained from biosensors. The goal of the study was to help choose alternative ways to build a system that can recognise emotions.

Similarly, a recent SLR conducted by Singh and Hamid~\cite{singh2022cognitive} presented the state-of-art computational methods and technologies aiding the automated detection of mental disorders. This study explored relevant literature between 2010 and 2021 with the \textit{Preferred Reporting Items for Systematic Reviews and Meta-Analyses} (PRISMA) as their method of review. The study recommended the need for multi-faceted approaches utilising data from physiological signals, behavioural patterns, and social media to efficiently and effectively detect the prevalence, type and severity of mental disorders. Furthermore, Garc{\'\i}a-Ponsoda's SMS work focused on the usage of feature engineering from EEG data and the application of AI algorithms to classify mental disorders~\cite{garcia2023feature}.

A secondary study on human status detection (HSD) was conducted by Sardar et al.,~\cite{sardar2022systematic}. Following the PRISMA approach, this study reviewed measures, tools, and machine-learning algorithms used in human status detection across 76 relevant studies. It offers valuable insights to researchers seeking HSD approaches employing various ML algorithms for their research.

All aforementioned secondary studies have explored machine learning applications in sentiment analysis, emotion detection, and mental disorder recognition, though none have focused specifically on burnout detection. Collectively, these studies highlight a strong foundation for applying ML techniques to emotion and mental health analysis, though a gap remains in their application to burnout detection specifically. 

Our SLR complements previous secondary studies by building on their foundational insights into sentiment analysis, emotion recognition, and mental disorder detection. While earlier reviews have mapped tools, data sources, and general ML applications across various domains, our study focuses on burnout-related indicators within software engineering, offering a performance-based review of machine learning models. Our work extends the contributions of prior research by connecting broader emotional and mental health detection efforts to the specific issue of burnout, thereby enriching the collective understanding of how ML can be applied in this critical area.

\section{Methodology}
\label{sec:_methodology}
In the systematic literature review conducted in this paper, we followed the well-defined guidelines for conducting an SLR~\cite{kitchenham2004procedures}: furthermore, we augmented the results of the relevant papers by adding one iteration of forward, and one of backward snowballing, as proposed in the guidelines of~\cite{wohlin2014guidelines}. 

The first author created a detailed review protocol, conducted searches, filtered the studies, and performed data extraction and synthesis. All of these activities were done under the supervision of the second and third authors. The third author conducted a supplementary review during data extraction.

\subsection{Research Questions}
To help the software engineering researchers develop machine learning-based models for detecting early symptoms of burnout based on the best knowledge across the previous studies, and to assist practitioners in making effective decisions on the array of machine learning models utilised for the early detection of burnout symptoms among software engineers, we define five research questions:

\textbf{RQ1: How have machine learning techniques been used, in past studies, for the early identification of burnout?}~
\\\textit{Rationale.} The goal of this question is to get a comprehensive view of the implementation of machine learning techniques to detect early signs of burnout in the context of SE. By acknowledging how machine learning was implemented in past studies, we can determine the direction of burnout research and the direction of future studies on this topic.

\textbf{RQ2:~What types of \textit{input} have been isolated in the past to investigate developers' behaviour, particularly related to their emotions, feelings, stress and relationships?}
\\\textit{Rationale}. Early symptoms of emotional exhaustion can be recognised by analysing human emotions, stress, and their relationships with other factors. Using machine learning techniques may help in detecting emotions, stress and inter-relationships automatically. In this question, we intend to acknowledge what common types of input have been used in detecting human emotions, feelings, stress and/or inter-relationships automatically. We also intend to obtain the dependent variables (i.e., the machine learning classes) implemented in the machine learning models of our primary studies. 

\textbf{RQ3:~Which \textit{ML-based modelling techniques} perform best in predicting the early signs of burnout in software engineers?} 
\\\textit{Rationale.} Developing machine learning models for the early detection of burnout has been an active field of research for the past decade. The empirical studies have individually reported their performance (in terms of precision, recall, F-score, and/or accuracy), but an overall comparison between ML approaches is still missing. In this question, we intend to acknowledge which ML method provided the best performance in the early detection of burnout.

\textbf{RQ4:~Which \textit{datasets} perform best when used in ML-based models at predicting the early signs of burnout in software engineers?}
\\\textit{Rationale.} Datasets are a crucial component in developing machine learning classifiers~\cite{paullada2021data,liebchen2008data,nitesh2019data}. Various types of datasets have been provided widely on software repositories. In this SLR, we focus on archival text communication types, such as issue reviews and/or discussions, bug reports, mailing lists, and online chat communication. In this question, we intend to present a synthesis of current knowledge on the impact of the datasets on model performances.

\subsection{Identification of Relevant Studies}
\begin{figure*}[t]
    \centering
    \includegraphics[scale=0.37]{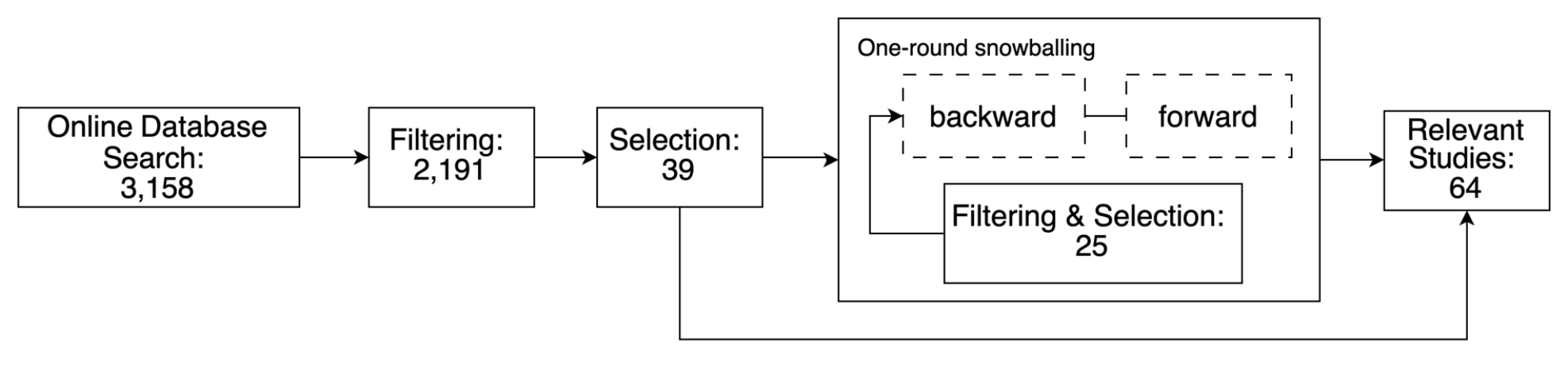}

    \caption{Pipeline used in this SLR to obtain the relevant primary studies.}
    \label{fig:_methodology}
\end{figure*}

The mechanism for identifying relevant studies is outlined in Figure~\ref{fig:_methodology}: in the section that follows, we detail its processes alongside the quantity of papers collected at each stage.

\subsubsection{Definition of the search string}
We defined our search query to search and identify relevant primary papers for this SLR. Our search query was obtained from the research questions via discussions.

Initially, we defined our search query with the terms 'automatic', 'detection', 'software engineering', 'software development', and 'burnout', with the Boolean operator AND and OR in between the terms. These terms were derived from the research questions. Below is the first search query:

\begin{mybox_old}
	 (automatic OR detection) AND (“software development” OR “software engineering”) AND burnout. 
\end{mybox_old}

We opted for the relevant papers as our control papers to initially check this query. We found that we still missed some other relevant papers, particularly papers that relate to factors (e.g. emotions, sentiments, and burnout) closely associated with burnout~\cite{zapf2002emotion,rey2016emotional,abellanoza2018burnout,spiller2021emotion}. Hence, we modified our search terms and employed additional terms: 'machine learning', 'sentiment', 'stress', and 'emotion' and utilised Boolean operator OR and AND. 
The final query in this study has then become: 

We employed the terms “burnout” and “software engineering” in the search query as these were the key terms of our research questions. Additionally, we used “software development” as a term to narrow the scope of a broader search of studies. Furthermore, using the Boolean operator OR to combine two terms “automatic” and “detection” to help in evaluating both RQ1 and RQ3.

\begin{mybox_new}
    (automatic OR detection OR "machine learning") AND 
    (“software development” OR “software engineering”) AND (burnout OR stress OR emotion OR sentiment). 
\end{mybox_new}


Using the “Full Text” and “Full text and Metadata” options, we utilised the search boxes given by each online library and conducted searches. 
We used the second string from the revised set of search queries as the final one in our search. 
The union of the results from each of these libraries generated 3158 documents from 2000 and 2025.



\subsubsection{Selection of Databases}
For a comprehensive search recommended in~\cite{Kitchenham07guidelinesfor}, we utilised several reliable online databases (e.g. IEEE Xplore Digital, ACM Digital Library,  Elsevier, and Springer). We expanded our search by utilising additional reputable online resources, including Google Scholar, Sage, PeerJ, Taylor \& Francis.

We used the group of online databases mentioned later for several reasons: 1) the databases contains journals in related fields such as information technology, computer science, information systems and so on; 2) Google Scholar was one of the online databases recommended in~\cite{brereton2007lessons} and used in a forward snowballing as recommended by Wohlin~\cite{wohlin2014guidelines}; 3)  PeerJ - an open academic journal - that is widely used in SE field; 4) PeerJ, Taylor and Francis, and Sage papers (e.g.~\cite{P4_destefanis2016software,P18Ostrovsky2012,P11van2018under}) were cited by papers published in IEEE, ACM, and Springer~\cite{lin2022opinion,yordanova2021agile}; 5) Additionally, it was explicitly suggested in~\cite{brereton2007lessons} to conduct a comprehensive search utilising a variety of databases.

 \subsubsection{Filtering and inclusion criteria} 
We manually collected the papers based on the inclusion and exclusion criteria summarised in Table~\ref{tbl:_in_ex_criteria}. The criteria table was derived from the research questions. We employed 4 inclusion criteria and 3 exclusion criteria

\begin{table}[htpb]
  \caption{Inclusion and Exclusion Criteria}
   \label{tbl:_in_ex_criteria}
   \begin{tabular}{cp{11cm}l}
    \toprule
    \multicolumn{2}{c}{Inclusion Criteria}\\
    \midrule
    IC1 & The paper must be a peer-reviewed paper and published in conferences,  workshops, or journals.  \\
IC2 & The paper must be accessible online  (i.e., PDF files available in the selected databases or through Google). \\
IC3 & The study must be published between the years 2000 and 2025. \\
IC4 & The study discussed or mentioned the early detection of burnout, or reported machine learning as its approach to predict burnout in software engineering.\\
    \midrule
    \multicolumn{2}{c}{Exclusion Criteria} \\ 
    \midrule
EC1 & The non-English paper contains non-English data 
source used.\\
EC2 & Publishers lack credibility. The reputable publishers include IEEE, ACM, Elsevier, Springer,  Sagepub, Taylor and Francis, and PeerJ.\\
EC3 & The paper is not a full research publication (e.g., abstract only submissions, doctoral symposium articles, presentations, tutorials, posters, forewords, survey).\\
  \bottomrule
\end{tabular}
\end{table}

We included peer-reviewed and accessible papers to achieve unbiased studies published at a workshop, conference, or journal (IC1 and IC2). We only looked at papers published between 2000 and 2025 (see IC3). IC4 required us to search for studies that mentioned and/or discussed burnout, specifically studies that used machine learning, not only in the title and abstract but also throughout the text (e.g., introduction/background, aims, methods, results, and discussions). 

We formulated our exclusion criteria to facilitate the elimination of any non-English and incomplete publication research (EC1 and EC3). Additionally, we exclude publications not published by IEEE, ACM, Elsevier, Sagepub, Taylor and Francis, or PeerJ (EC2).

As a result of applying our inclusion and exclusion criteria during the filtering step, 2,191 papers were discovered to be relevant for the initial batch, which used as input for the selection phase.

During the selection stage, we thoroughly scrutinised the papers. The examination began with each study's title. We removed the papers containing a secondary study title. If the paper's title explicitly mentioned the term \textit{‘burnout’}, \textit{‘stress’},  \textit{‘emotion’},  and/or, \textit{sentiment,} the first author further examined the abstract to get more detailed information. If the abstract indicated a study on burnout or early burnout detection, the paper proceeded to the next phase. If the term \textit{‘burnout’}, \textit{‘stress’},  \textit{‘emotion’}, and/or \textit{sentiment} were not present in the title, the author read the abstract; if it did not provide sufficient information, the author skimmed the paper for information such as the introduction/background, the purpose, the method, the results, and the conclusion that would indicate the paper's focus on burnout. The second author re-checked or independently analysed the results by randomly selecting papers and examining them using the same procedures used by the first author. 
We resolved any disagreements by having detailed discussions to review and reconcile the results. This process involved examining the discrepancies closely and reaching a consensus through thorough deliberation.

After filtering the relevant papers (a total of 39), we reexamined each one according to Wohlin~\cite{wohlin2014guidelines} to decide if it should be included in the snowballing phase. The reexamination aimed to ensure that the collected papers meet the characteristics of the start set well~\cite{wohlin2014guidelines}.

\subsubsection{Snowballing}
After obtaining 39 relevant papers, we conducted backward and forward snowballing as recommended in the guidelines of ~\cite{wohlin2014guidelines}. These 39 papers were considered as the start set. 

\textit{Backward Snowballing} - In this step, we adopted the backward approach suggested by Wohlin~\cite{wohlin2014guidelines}: for the initial iteration of backward snowballing, we examined the bibliographies of the papers obtained in the preceding phase. If the reference was already in the start set, it was disregarded; otherwise, similarly to the previous step (“Filtering and inclusion criteria”), the title, abstract, and relevant sections of the candidate paper were examined to determine whether the new paper should be included in the next phase or not. 

\textit{Forward Snowballing} - Based on the citations to each paper in the start set, we determined the candidate papers in this phase. According to Wohlin~\cite{wohlin2014guidelines}, we used Google Scholar to identify academic papers that cite the core papers. Then, we examined each paper that cites the paper in the start set by determining whether or not the candidate paper was already in the start set. If the paper was in the start set, it was disregarded; if it was not in the start set, a similar method was followed during the “selecting paper” step.

\textit{Final set} - We found 25 new papers with the snowballing method. The total outcome of the SLR, we reviewed 39+25= 64 papers published between 2000 and 2025. The complete pipeline of how we obtain our primary studies is depicted in Figure~\ref{fig:_methodology}. 

We did not apply an interrater to evaluate our results; however, as recommended in Kitchenham's guidelines~\cite{Kitchenham07guidelinesfor} a test-retest also can be used to check the reliability of the inclusion decisions during the selection process. Therefore, the first author conducted the filtering of the final set, utilising the inclusion and exclusion criteria on separate occasions and re-evaluated the sample of our primary studies after the screening process in order to examine the consistency of our inclusion/exclusion criteria. Moreover, all included and excluded papers were discussed with the second or third author. 
Additionally, if there were any disagreements about the papers, we resolved them by having detailed discussions to review and reconcile the results.

\subsection{Study Quality Assessment}
We applied the quality assessment procedures recommended by Kitchenham and Charter~\cite{kitchenham2004procedures} to identify papers that can address the research questions described above. We created an assessment criteria form, as shown in Table~\ref{tbl:_assessment}, and employed the questions on it to evaluate each of the studies. We identified: 

\begin{table*}[htpb!]
\centering
\caption{Assessment Criteria Form}
\label{tbl:_assessment}
\begin{tabular}{lp{8cm}l}\hline
No. & Questions & Focus\\
\hline
1  & Are the aims clearly stated? & RQ1 \\
2  & Does the paper clearly discuss the findings explaining the possibility of experiencing the risk of burnout or negative behaviour, or mention burnout or negative behaviour (e.g. negative emotion, feeling, stress or relationship)? &  RQ1 \\
3  & Does the paper utilised the machine learning approach(s)? & RQ1- RQ4  \\
4  & Does the paper report the performance of the machine learning models? & RQ3 \\
5  & Does the paper clearly state the datasets employed? & RQ4  \\\hline
\end{tabular}
\end{table*}

 (1) the objective of each study. The study's clearly stated purpose may provide us with broad insights into answering RQ1;

 (2) whether any data described in the papers are associated with burnout risk or negative behaviour (e.g., negative emotion, feeling, stress or relationship). The analysis of this data will address RQ1;

 (3) whether any machine learning  method(s) or approach(es) are utilised in the papers. The papers should state the machine learning employed and describe the model built. This question will address RQ1-RQ4;

 (4) whether performances of the machine learning models are reported in the papers. The papers should report the performance of the models (e.g. precision, recall, f-measure, or accuracy). This question will address RQ3;

 (5) whether any datasets utilised are reported. The paper should state datasets used along with the size of datasets employed in developing stage of models. This question will address RQ4.

\subsection{Data Extraction}
\label{sec:_meth-extraction}
After thoroughly examining the whole text of each primary study, we extracted its data. Both qualitative and quantitative data of the studies were extracted: the qualitative data included (i) the aim of the study, (ii) the variables or features, (iii) the data sources or datasets, (iv) the methods and (v) the results. The quantitative data included the sample size or the number of participants; the number of data sources, and the size of datasets, the performance measures (e.g. f-measure, precision, recall, and accuracy). 

The extracted data was compiled into a shared spreadsheet\footnote{The worksheet is available at \url{https://github.com/phd-work-22/SLR-Early-Identification-of-Burnout}}.
The first author executed the data extraction process; the second author double-checked the results by selecting studies at random and comparing them to the first author's to assess their consistency. 
If the disagreements existed, we further discussed about the conflicted papers. 

\subsection{Data Synthesis}
\label{sec:_meth-synthesis}
We integrated both qualitative and quantitative data from our worksheet to collate and synthesise all of the studies. We synthesised our data in three rounds, and several iterations. 

\textbf{Round One (${R_1}$) -- }  In the first round, we discussed the main characteristics of the studies that focused on the early detection of burnout. This round was conducted in two iterations  ($R_1I_1$ and $R_1I_2$): 

\begin{itemize}

\item \textbf{${R_1I_1}$ -- } In the first iteration, we discussed all the 64 studies focusing on the early detection of burnout. We then collated the following data from the resulting papers: title, aim of the paper, authors, year of publication, machine learning method (s) included and results. 

\item \textbf{${R_1I_2}$ -- } In the second iteration, the first and third authors discussed and summarised the studies based on the extracted data. The draft of the synthesis was discussed among all the authors. Any disagreement about the draft was resolved via discussions.
\end{itemize}

\textbf{Round Two (${R_2}$) -- }  During the second round, we discussed the types of classification employed in the 64 primary studies. This round was conducted in three iterations ($R_2I_1$, $R_2I_2$ and $R_2I_3$):

\begin{itemize}

\item \textbf{${R_2I_1}$ -- } The first author extracted the relevant results (e.g. title, data source, variables or features, classes, methods used) from the papers. Following the analysis of the results and using an open sorting approach~\cite{upchurch2001using}, the first author grouped studies into five different types of models: i) studies classifying comments/text into sentiment polarity; ii) studies classifying comments into emotion classes; iii) studies using sensor data and classifying them into emotion classes; iv) studies classifying text into toxic/non-toxic classes; v) studies predicting attrition.  All of these subgroups were made available to the other authors, and its synthesis was directly discussed with the second author. 

\item \textbf{${R_2I_2}$ -- }  The first author represented the revised spreadsheets agreed upon by consensus in the previous iteration. Based on the spreadsheets, we discussed three types of classification: i) emotion and stress detection, ii) attrition prediction, and iii) toxic detection. Then, we identified 5 (five) kinds of independent variables:  text, sensor, utterances, movement, and facial expression. The first author further identified the dataset(s) used in each study and put them into a shared table. This dataset table was discussed with the second author. The consensus of the table was reached by the end of the iteration via discussions.

\item \textbf{${R_2I_3}$ -- } The first author gathered all the spreadsheets produced from the previous iterations and visualised them into different graphs.
\end{itemize}

Any disagreements during this round were resolved via discussions.  Researcher bias in defining the types of classification, kinds of independent variables, and names of datasets were reduced by the iterations of this phase and with the discussions between the first and the second author. 

\textbf{Round Three (${R_3}$) --} In the third round, we discussed the performances of models developed in each study. This round was conducted in three iterations to reduce research bias during this round  ($R_3I_1$, $R_3I_2$ and $R_3I_3$). 

\begin{itemize}

\item \textbf{${R_3I_1}$ -- } The first author collated the performances reported in each of the 64 studies and summarised them in a spreadsheet. The extracted data used for the report was: i) precision, ii) recall, iii) f-score, and iv) accuracy of the ML models. All the authors then discussed and decided which studies would be included in the synthesis as only 57 studies reported the complete performances, in terms of the 4 attributes selected. After discussions between each other, we decided to collate only those 57 `complete' studies for our synthesis.  

\item \textbf{${R_3I_2}$ -- } The first author analysed all data performances and put them into tables in a spreadsheet. The three authors then discussed the performance data, in particular focusing on missing data on performance in the studies. At the end of our discussion, we agreed to put all the performances (e.g. precision, recall, f-measure, and accuracy) in all the performance graphs. We also decided to use the R tool to create our graphs.

\item \textbf{${R_3I_3}$ -- } Initially, we created several boxplot graphs by studies and methods. As there were any disagreements, we discussed the graphs in particular categorising each performance model across the studies. The first author then revised the graphs and presented them to the second and third authors. Initially, we determined four types of model performance graphs: ‘Emotion Detection’, ‘Emotion Detection with sensor data’, ‘Toxic Detection’, ‘Attrition Prediction’, and ‘Model performances by Datasets’. Although we differentiate sensor and movement as different type of input, after discussion we decided to merge these type of inputs into one graph. Therefore, we ended up the graphs of the model performance: ‘Model performance of emotional and stress detection’, ‘Model performances of emotion and stress detection with sensors and movement’, ‘Model performances of attrition detection’, and ‘Model Performances of toxicity detection’. Any disagreement about the figures was resolved through discussion between the first two authors.

\item \textbf{${R_3I_4}$ -- } The first author made the synthesis report of the graphs. The report was then discussed with the second and third authors. Any feedback and suggestions from the second and third authors were discussed and changes were made after the consensus.
\end{itemize}

All the collation and synthesis in the three rounds were done iteratively: these syntheses took an overall 8 weeks. The discussions were conducted in both online and in-person meetings.

\label{sec:_findings}
\begin{table*}[htpb!]
\footnotesize
\centering
\caption{Table of analysed primary studies. Studies have employed more than one machine learning algorithm. Studies with bold fonts reported the performances in figures only or did not report the performances in their papers.}
\label{tbl:_table_synthesis}
\begin{tabular}{p{3cm}p{3.5cm}p{4.5cm}}
\hline
Categories & 
Paper ID & Machine Learning Techniques
\\\hline  
    C1 Emotional Dimension as Burnout Predictors & 
        \cite{S1_Mantyla2016} &
        Linear Regression~\cite{S1_Mantyla2016}, Zero R~\cite{S1_Mantyla2016}\\
    C2 Burnout Prediction & \cite{S92_nath2021burnoutwords}, \textbf{\cite{S24_Dovleac2021}} & SVM~\cite{S92_nath2021burnoutwords}, ANN~\cite{S24_Dovleac2021}
    \\
    C3 Emotion. Dimension and Stress Detection & 
        \cite{S8_GirardiDaniela2020,S14_novielli2022sensor,S42_Nogueira2013,S46_GirardiLanubile2021,S77_vrzakova2020affect,S79_fritz2016leveraging,S82_nogueira2015modelling,S83_naegelin2023interpretable,S84_epp2011identifying,S21_srikanteswara2024machine,S24_Dovleac2021,S30_gamage2022machine,S44_androutsou2023automated,S52_reddy2018machine,S74_booth2022toward,S75_padha2022quantum,S88_pepa2020stress,S89_alberdi2018using,S93_vizer2009automated,S2_Gachechiladze2017,S4_murgia2018exploratory,S19_imran2024emotion,S27_ballesteros2024facial,S33_bleyl2022emotion,S34_maheshwarkar2021analysis,S38_Muller2015,S48_islam2019marvalous,S63_cabrera2020classifying,S68_wagan2025multilabeled,S72_yang2021behavioral,S20_singh2024softment,S12_soto2021observing,S81_anany2019influence,S91_klunder2020identifying,S6_Islam2018,S23_manikandan2024stress,S29_jayathilake2023accurate,S87_carneiro2012multimodal,S78_rissler2020or}, \textbf{\cite{S23_manikandan2024stress,S26_geeth2024identification,S28_awan2023creating,S43_radevski2015real,S55_kolakowska2013emotion,S87_carneiro2012multimodal}} &
        
        AB~\cite{S44_androutsou2023automated,S78_rissler2020or}, 
        ANN~\cite{S21_srikanteswara2024machine},
        BERT~\cite{S33_bleyl2022emotion,S34_maheshwarkar2021analysis,S20_singh2024softment,S19_imran2024emotion}
        NN~\cite{S82_nogueira2015modelling,S55_kolakowska2013emotion}, 
        DT~\cite{S42_Nogueira2013,S48_islam2019marvalous,S79_fritz2016leveraging,S82_nogueira2015modelling,S84_epp2011identifying,S21_srikanteswara2024machine,S29_jayathilake2023accurate,S30_gamage2022machine,S52_reddy2018machine,S78_rissler2020or,S38_Muller2015,S55_kolakowska2013emotion}, 
        GBT~\cite{S48_islam2019marvalous}, 
        J48~\cite{S8_GirardiDaniela2020,S2_Gachechiladze2017}, KNN~\cite{S8_GirardiDaniela2020,S46_GirardiLanubile2021,S21_srikanteswara2024machine,S44_androutsou2023automated,S52_reddy2018machine,S75_padha2022quantum,S78_rissler2020or,S26_geeth2024identification,S55_kolakowska2013emotion,S87_carneiro2012multimodal}, 
        MLP~\cite{S8_GirardiDaniela2020,S48_islam2019marvalous,S21_srikanteswara2024machine,S74_booth2022toward}, 
        NB~\cite{S8_GirardiDaniela2020,S2_Gachechiladze2017,S79_fritz2016leveraging,S48_islam2019marvalous,S21_srikanteswara2024machine,S30_gamage2022machine,S78_rissler2020or,S2_Gachechiladze2017,S34_maheshwarkar2021analysis,S55_kolakowska2013emotion},  
        RF~\cite{S8_GirardiDaniela2020,S12_soto2021observing,S14_novielli2022sensor,S77_vrzakova2020affect,S46_GirardiLanubile2021,S42_Nogueira2013,S82_nogueira2015modelling,S83_naegelin2023interpretable,S21_srikanteswara2024machine,S30_gamage2022machine,S34_maheshwarkar2021analysis,S44_androutsou2023automated,S52_reddy2018machine,S74_booth2022toward,S81_anany2019influence,S78_rissler2020or},  
        SLP~\cite{S42_Nogueira2013,S78_rissler2020or}, 
        SVM~\cite{S8_GirardiDaniela2020,S14_novielli2022sensor,S42_Nogueira2013,S2_Gachechiladze2017,S79_fritz2016leveraging,S82_nogueira2015modelling,S83_naegelin2023interpretable,S21_srikanteswara2024machine,S44_androutsou2023automated,S75_padha2022quantum,S78_rissler2020or,S2_Gachechiladze2017,S4_murgia2018exploratory,S26_geeth2024identification,S48_islam2019marvalous,S55_kolakowska2013emotion,S23_manikandan2024stress} ,
        GBM~\cite{S83_naegelin2023interpretable,S44_androutsou2023automated},
        Logistic Regression~\cite{S21_srikanteswara2024machine,S30_gamage2022machine,S52_reddy2018machine,S72_yang2021behavioral},
        Linear Regression~\cite{S21_srikanteswara2024machine},
        LSTM~\cite{S26_geeth2024identification,S34_maheshwarkar2021analysis,S72_yang2021behavioral},        CNN~\cite{S23_manikandan2024stress,S26_geeth2024identification,S27_ballesteros2024facial,S34_maheshwarkar2021analysis,S68_wagan2025multilabeled},
        RNN~\cite{S24_Dovleac2021,S26_geeth2024identification,S23_manikandan2024stress},
        XGBoost~\cite{S30_gamage2022machine,S44_androutsou2023automated},
        CatBoost~\cite{S30_gamage2022machine},
        RUSBoost~\cite{S44_androutsou2023automated},
        ZeroR~\cite{S78_rissler2020or},    
        Voting Classifier(SVM, NB, and RF)~\cite{S91_klunder2020identifying}
         \\
    C4 Sentiment and Emotions & 
        
        \cite{S31_silva2023using}, \textbf{\cite{S39_Pletea2014,S40_ortu2016arsonists} } &
        
        Hierarchical Classifier~\cite{S39_Pletea2014}, 
        NB~\cite{S39_Pletea2014}, 
        RF~\cite{S81_anany2019influence},
        SentiStrength~\cite{S31_silva2023using,S40_ortu2016arsonists} \\

    C5 Attrition prediction & \cite{S17_trinkenreich2024predicting,S22_ozakca2024artificial,S37_Garcia2013} & 
    AdaBoost~\cite{S22_ozakca2024artificial},
    Bayesian Classfr.~\cite{S37_Garcia2013}
    DT~\cite{S17_trinkenreich2024predicting,S22_ozakca2024artificial},
    KNN~\cite{S22_ozakca2024artificial},
    Logistic Regression~\cite{S22_ozakca2024artificial},
    NB~\cite{S22_ozakca2024artificial},
    RF~\cite{S17_trinkenreich2024predicting,S22_ozakca2024artificial},
    SVM~\cite{S22_ozakca2024artificial} \\
    
    C6 Detecting Unhealthy Relationships as an Early Warning to Prevent Burnout & 
        \cite{S7_Raman2020,S9_Cheriyan2021,S13_qiu2022detecting,S16_Sarker2023,S18_ferreira2024incivility,S47_sarker2022identification,S60_mishra2024exploring,S70_Sarker2020,S71_bhat2021say,S73_rahman2024words} &

        CART~\cite{S18_ferreira2024incivility},
        ChatGPT~\cite{S60_mishra2024exploring,S73_rahman2024words}
        CNN~\cite{S70_Sarker2020,S47_sarker2022identification,S7_Raman2020}, 
        DNN~\cite{S15_Sarker2023automated}, 
        DT~\cite{S15_Sarker2023automated,S47_sarker2022identification}, 
        GBT~\cite{S15_Sarker2023automated,S47_sarker2022identification}, 
        KNN~\cite{S18_ferreira2024incivility},
        LR~\cite{S15_Sarker2023automated}, 
        Lexicon-based~\cite{S16_Sarker2023},
        LR~\cite{S18_ferreira2024incivility,S47_sarker2022identification},
        NB~\cite{S18_ferreira2024incivility},
        RF\cite{S9_Cheriyan2021,S15_Sarker2023automated,S18_ferreira2024incivility,S47_sarker2022identification,S13_qiu2022detecting}, 
        SVM~\cite{S7_Raman2020,S70_Sarker2020,S9_Cheriyan2021,S15_Sarker2023automated,S18_ferreira2024incivility,S47_sarker2022identification}, 
        BERT~\cite{S70_Sarker2020,S9_Cheriyan2021,S15_Sarker2023automated,S16_Sarker2023,S41_Sinha2016,S47_sarker2022identification,S7_Raman2020,S71_bhat2021say,S73_rahman2024words},
        LSTM~\cite{S15_Sarker2023automated,S47_sarker2022identification}\\
\hline
\end{tabular}
\end{table*}

\section{Findings - (RQ1)}
\label{sec:_findings_rq1_early}
In this section we show the findings to address RQ1: for this purpose, as shown in Table~\ref{tbl:_table_synthesis}, we group studies into six categories. As our SLR study only focuses on the papers that employ machine learning techniques, we analyse six categories that present ML-based studies.  
In this section, we scrutinise our analysis of machine learning classifiers employed by the studies. 

\subsection*{[C1] Emotional Dimension as Burnout Predictors}
In this category, researchers focused their studies on the impact of emotions on burnout. For instance, M\"{a}ntyl\"{a} et al.,~\cite{S1_Mantyla2016} investigated three common emotional characteristics (Valence, Arousal and Dominance - VAD) and used Linear Regression and Zero R algorithms, based on problem reports of SE projects, to identify productivity loss and burnout symptoms. They claimed that higher levels of valence and lower levels of arousal may lessen burnout, especially among highly experienced developers. In addition, they provided a common starting point that may be used to predict burnout in SE. 

It should be noted that the aforementioned study utilised just comments or interview transcription as their major data source; the results of this study may be limited, but it might be expanded to incorporate other variables besides text, and used in prediction models or classifiers. 

\subsection*{[C2] Burnout Prediction}
In this category, two studies attempted to predict burnout by considering text as well as heart's pulse and oxygen level to measure the level of burnout or to detect the existence of potential burnout~\cite{S92_nath2021burnoutwords,S24_Dovleac2021}. One work done by Nath and Kurpicz-Briki~\cite{S92_nath2021burnoutwords} built the burnout datasets containing words retrieved from the transcript of patients experiencing burnout. They employed this dataset and used an SVM classifier to label whether the text contained burnout or not. Their findings reported their model's accuracy over 0.75. Meanwhile, another work~\cite{S24_Dovleac2021} built a wearable sensor that may measure the level of burnout by considering the heart rate and oxygen level of the user. The model of the work was trained using data from 30 subjects over one month, with oversight from a psychologist to ensure consistency. It was then tested on 5 subjects: one had a burnout level over 75\%, one was between 50–75\%, and the others were below 50\%.

\subsection*{\textbf{[C3]} Emotion and Stress Detection}
Some studies developed machine-learning-based classifiers to detect `arousal' and `valence' using psycho-physiological variables as input~\cite{S8_GirardiDaniela2020,S14_novielli2022sensor,S42_Nogueira2013,S46_GirardiLanubile2021,S77_vrzakova2020affect,S79_fritz2016leveraging,S82_nogueira2015modelling}. These studies implemented Naïve Bayes (NB), K-Nearest Neighbour (KNN), J48, Support Vector Machine (SVM), Random Forest (RF), and Single-Layer Perceptron (SLP) in their models. Two studies~\cite{S8_GirardiDaniela2020,S46_GirardiLanubile2021} not only consider the psycho-physiological variables but also self-report text to measure the emotional dimensions (e.g., valence and arousal). Studies found that developers experienced a variety of emotions while completing their programming-related duties~\cite{S8_GirardiDaniela2020,S79_fritz2016leveraging}. In~\cite{S8_GirardiDaniela2020}, the authors discovered that `valence' was positively connected with perceived progress. Another study found that positive long-term affect was associated with after-task valence, indicating that prior well-being influences happiness after the code review task~\cite{S77_vrzakova2020affect}. 

~
Many studies developed machine learning models to identify a range of emotions such as anger, sadness, fear, surprise, joy, happiness, and love, in addition to detecting VAD (Valence, Arousal, and Dominance) attributes. For example, Murgia et al.~\cite{S4_murgia2018exploratory} created an SVM-based classifier that identifies emotions in comments, with human annotators validating the labels. Gachechiladze et al.~\cite{S2_Gachechiladze2017} focused on detecting self-directed anger to help prevent burnout, using SVM, J48, and Naïve Bayes algorithms. Other research utilised a variety of machine learning techniques, including K-nearest neighbours (KNN), SVM, neural networks (NN), random forests (RF), and BERT-based models. A summary of these studies and their classifiers is presented in Table \ref{tbl:_table_synthesis}.

Some studies built machine-learning classifiers to detect and predict stress, utilising advanced algorithms like Random Forest (RF), Support Vector Machine (SVM), Decision Trees (DT), Neural Networks (NN), and others, as noted in Table~\ref{tbl:_table_synthesis}. 

While some research focuses solely on stress detection~\cite{S23_manikandan2024stress,S29_jayathilake2023accurate}, others examine related symptoms like depression, anxiety, and positive emotions such as excitement and relaxation~\cite{S6_Islam2018,S21_srikanteswara2024machine,S30_gamage2022machine}. A prototype tool has been developed to detect various emotional states, including stress and relaxation, by measuring valence and arousal~\cite{S6_Islam2018}. 

Additionally, studies have explored the impact of various factors on mental distress in IT workers~\cite{S21_srikanteswara2024machine,S29_jayathilake2023accurate}. Some have used classifiers like SVM, RF, and Gradient Boosting Machine to assess both emotional dimensions and stress~\cite{S83_naegelin2023interpretable}, while others have proposed tree-based classifiers for detecting stress-related symptoms~\cite{S84_epp2011identifying}.

Upon our findings, machine learning methods such as DT, KNN, NB, RF, and SVM are the popular methods used among the studies; more than five studies employed these approaches. 

\subsection*{\textbf{[C4]} Sentiment and Emotions}
This category contains various studies that have proposed sentiment-based classifiers to analyse and classify software developers' emotions. We consider these in our analysis because they employed sentiment tools, which were optimised by machine learning approaches~\cite{thelwall2010sentiment}, intended to investigate emotion.

 The work done by Silva et al.,~\cite{S31_silva2023using} classified the emotional polarity of developers by investigating the developers' personal tweets that were posted during working hours. This study employed SentiStrength to classify the posts. In addition, Pletea et al.,~\cite{S39_Pletea2014} used NLTK tool trained with Naive Bayes and Hierarchical classifiers to classify comments into negative or positive comments. In this study, security-related comments contained more negative emotional responses than non-security-related comments. A study investigated the behaviour shown by the agile developers. This work evaluated the developer's behaviour (e.g., politeness, sentiment, and emotions) by employing the Sentistrength tool~\cite{S40_ortu2016arsonists}. 

All of the aforementioned findings indicate that, whether identifying or categorising emotions using a sentiment analysis approach (e.g. SentiStrength), researchers have relied mostly on developer comments or texts.

\subsection*{\textbf{[C5]} Attrition Prediction}
This category of study focuses on the turnover prediction as a potential consequence of burnout. For example, the earliest study focused on the intention to leave the projects was conducted by Garcia et al.~\cite{S37_Garcia2013}. Their work built Bayesian classifiers that may predict the likelihood of a developer (a contributor) becoming inactive. They reported that emotional expression is a significant indicator of a contributor's likelihood to remain active in the project. A recent study in~\cite{S17_trinkenreich2024predicting} investigated the relationship between the engagement, opportunities to learn and the employee's turnover. They built machine-learning classifiers with RF and DT as their algorithms and may predict the employee's turnover with accuracies and F-Scores above 80\%. Within the same year, a work reported in~\cite{S22_ozakca2024artificial} that their machine learning classifier can predict employee turnover with high accuracy, in which RF and AB algorithms performed the highest performance above 90\%. They also reported that company culture, compensation, and employee-manager relationships are crucial factors influencing employee loyalty and reducing attrition. 

\subsection*{\textbf{[C6]} Detecting Unhealthy Relationships as an Early Warning to Prevent Burnout} 
This type of study focuses on interpersonal interactions as the primary predictor for (early) burnout detection. The research conducted by Raman et al.,~\cite{S7_Raman2020} suggested an alternative method for minimising stress and burnout among software engineers. Instead of detecting emotions in textual communication, they have recommended classifying it as toxic and non-toxic language based on its content. Following this work, further studies addressed a similar topic as described in Table~\ref{tbl:_table_synthesis}.  Several classical ML algorithms were implemented (e.g. SVM, GBT, DT, RF), Deep Learning, and Transformer-based Model (e.g., BERT, Roberta, DistilBERT, ALBERT, XLnet).

\subsection*{\textbf{Summary of findings in RQ1}} Although using different types of input and datasets, we found the primary studies utilising machine learning approaches for four main study \textit{purposes}:
\begin{itemize}
    \item detecting emotions, stress, and sentiment (subcategories [C1], [C3], and [C4])
    \item detecting burnout (subcategory [C2])
    \item predicting attrition (subcategory [C5])
    \item detecting toxic relationships (subcategory [C6])
\end{itemize}

In the rest of the paper, we will use these study \textit{purposes} to report on the types of input used to instrument ML approaches (Section~\ref{sec:_findings_rq2_types_of_ml}); the performances of the ML approaches by input (Section~\ref{sec:_findings_rq3_methods}); and the performances of the ML approaches by dataset (Section~\ref{sec:_findings_rq5_datasets}).

\section{Findings -  (RQ2)}
\label{sec:_findings_rq2_types_of_ml}

Overall, we found 64 studies that implemented machine learning methods to detect human behaviour. Since RQ2 is focused on the \textit{types of input} used in the analysed papers, we report below five types that were detected in the analysed studies:

\begin{enumerate}
    \item Text-based input, 
    \item Sensors-based, 
    \item Movement-based,
    \item Utterances-based, 
    \item Facial Expressions.
\end{enumerate}

\begin{figure*}[ht!]
    \centering
    \includegraphics[scale=0.55]{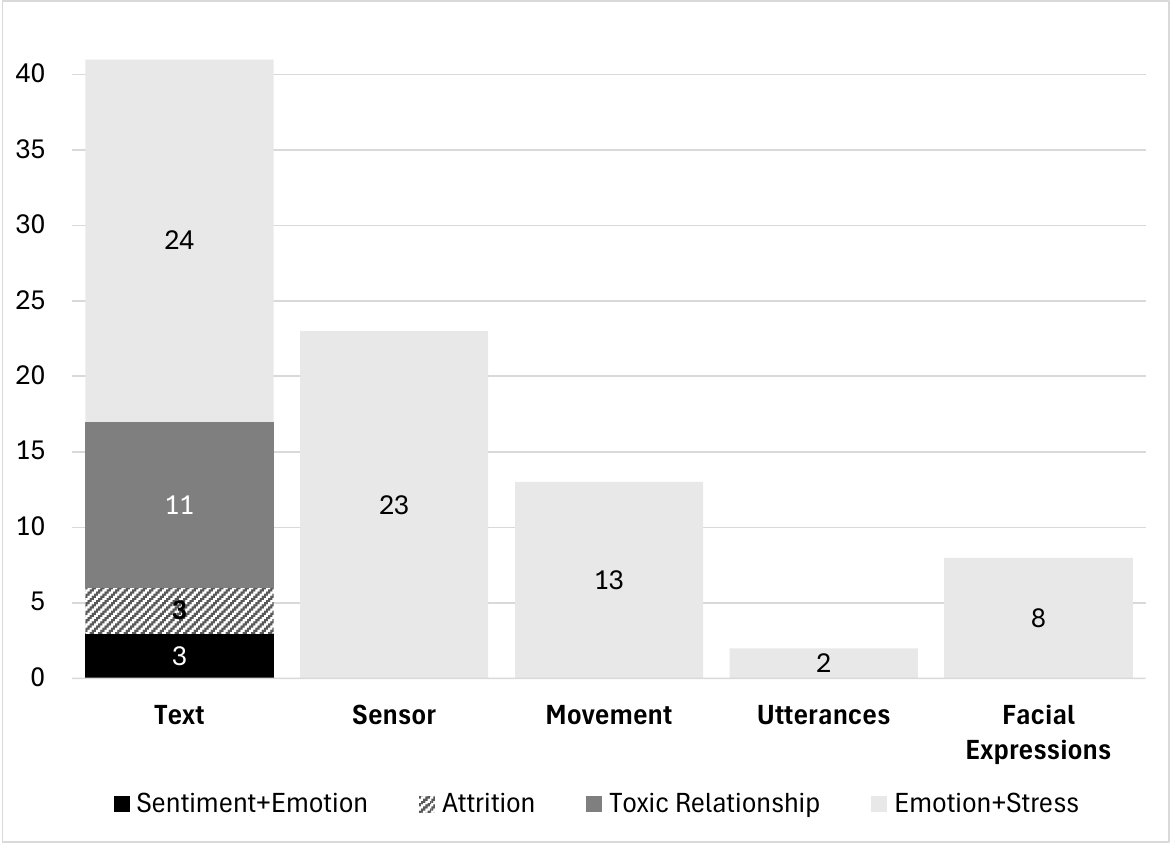}
    \caption{Number of papers grouped by purpose of study and input type.}
    \label{fig:_independent_vars}
\end{figure*}

It should be noted that 16 studies employed more than one type of input. Furthermore, we detected four main types of classification: a) Sentiment+Emotion b) Attrition, c) Toxic Relationship, and d) Emotion+Stress. Figure~\ref{fig:_independent_vars} shows the four main types of classification grouped by five input types (e.g., Text, Sensor, Movement, Utterances, and Facial Expressions). The analysis revealed that text-based inputs were the most commonly used across types of classification, with a significant number focusing on detecting emotions, stress, and toxic relationships. Other types of inputs (such as sensor-based data, movement, utterances and facial expressions), were less frequently employed but were also explored in specific contexts to enhance the detection of emotional and stress-related patterns.

\begin{table*}[htpb!]
\centering
\footnotesize
\caption{The list of studies grouped by their types of input and classification. All studies that employ sensors, movement, utterances, and facial expressions as their input are studies investigating emotions and stress detection. Bold studies use multiple input types, while regular studies use just one.}
\label{tbl:_types_of_input}
\begin{tabular}{p{2.5cm}p{1.5cm}p{7cm}}
    \hline
        \textbf & \# of studies & studies\\
    \hline
        \textbf{Text-based} &  & \\
          & 3 & \cite{S31_silva2023using,S39_Pletea2014,S40_ortu2016arsonists}\\
         & 24 & \cite{S1_Mantyla2016,S2_Gachechiladze2017,S4_murgia2018exploratory,S6_Islam2018,S19_imran2024emotion,S20_singh2024softment,S21_srikanteswara2024machine,S26_geeth2024identification,S30_gamage2022machine,S33_bleyl2022emotion,S34_maheshwarkar2021analysis,S48_islam2019marvalous,S52_reddy2018machine,S63_cabrera2020classifying,S68_wagan2025multilabeled,S85_munoz2022text,S91_klunder2020identifying,S92_nath2021burnoutwords,S93_vizer2009automated}, \textbf{\cite{S8_GirardiDaniela2020,S46_GirardiLanubile2021,S38_Muller2015,S44_androutsou2023automated,S72_yang2021behavioral}}  \\
         & 11 & \cite{S7_Raman2020,S9_Cheriyan2021,S13_qiu2022detecting,S15_Sarker2023automated,S16_Sarker2023,S18_ferreira2024incivility,S47_sarker2022identification,S60_mishra2024exploring,S70_Sarker2020,S71_bhat2021say,S73_rahman2024words}  \\
         & 3 & \cite{S17_trinkenreich2024predicting,S22_ozakca2024artificial,S37_Garcia2013} \\ 
        \hline
\textbf{Sensor-based} & & \\
        Skin conductance & 14 & \textbf{\cite{S8_GirardiDaniela2020,S12_soto2021observing,S14_novielli2022sensor,S38_Muller2015,S42_Nogueira2013,S44_androutsou2023automated,S46_GirardiLanubile2021,S55_kolakowska2013emotion,S72_yang2021behavioral,S77_vrzakova2020affect,S79_fritz2016leveraging,S82_nogueira2015modelling,S86_koldijk2016detecting,S89_alberdi2018using}}\\
        Heart-related sensors & 18 & \cite{S75_padha2022quantum,S78_rissler2020or}, \textbf{\cite{S8_GirardiDaniela2020,S12_soto2021observing,S14_novielli2022sensor,S24_Dovleac2021,S29_jayathilake2023accurate,S44_androutsou2023automated,S42_Nogueira2013,S46_GirardiLanubile2021,S55_kolakowska2013emotion,S72_yang2021behavioral,S74_booth2022toward,S79_fritz2016leveraging,S82_nogueira2015modelling,S83_naegelin2023interpretable,S86_koldijk2016detecting,S89_alberdi2018using}}
        \\ 
        Muscle and nerve signals & 3 & \textbf{\cite{S42_Nogueira2013,S55_kolakowska2013emotion,S82_nogueira2015modelling}} \\
        Oxygen rate & 5 & \textbf{\cite{S24_Dovleac2021,S29_jayathilake2023accurate,S44_androutsou2023automated,S72_yang2021behavioral,S82_nogueira2015modelling}}\\
        Respiratory signal & 1 & \textbf{\cite{S55_kolakowska2013emotion}}\\
        Neural signals & 4 & \cite{S43_radevski2015real},\textbf{\cite{S8_GirardiDaniela2020,S38_Muller2015,S79_fritz2016leveraging}}
        \\
        \hline
            \textbf{Movement-based} & & \\
         & 13 & \cite{S84_epp2011identifying,S87_carneiro2012multimodal,S88_pepa2020stress}, \textbf{\cite{S12_soto2021observing,S29_jayathilake2023accurate,S44_androutsou2023automated,S55_kolakowska2013emotion,S74_booth2022toward,S75_padha2022quantum,S77_vrzakova2020affect,S83_naegelin2023interpretable,S81_anany2019influence,S89_alberdi2018using}}\\
        \hline
            \textbf{Utterances} & & \\
         & 2 & \textbf{\cite{S28_awan2023creating,S72_yang2021behavioral}}\\
        \hline
            \textbf{Facial expressions} & & \\
         & 8 & \cite{S23_manikandan2024stress,S27_ballesteros2024facial}, \textbf{\cite{S14_novielli2022sensor,S28_awan2023creating,S55_kolakowska2013emotion,S75_padha2022quantum,S81_anany2019influence,S89_alberdi2018using}}\\       
    \hline
\end{tabular} 
\end{table*}

\textbf{Feature Extraction} -- In the studies applying text as their type of feature, the feature extraction methods vary from a single technique to combined techniques. Single techniques include applying bag-of-words (BoW) combined with unique uni- or bi-grams~\cite{S4_murgia2018exploratory}; using a single tool such as Google's perspective API~\cite{S7_Raman2020}; performing valence-arousal-dominance (aka VAD) calculations~\cite{S1_Mantyla2016,S8_GirardiDaniela2020,S46_GirardiLanubile2021,S92_nath2021burnoutwords}; using a combination of SentiStrength-SE tool, Software Engineering Arousal and Affective Norms for English Words Dictionary~\cite{S6_Islam2018} for calculating the valence and arousal score, the combination of lexical-, keyword-, and semantic-based features~\cite{S16_Sarker2023,S15_Sarker2023automated}, the combination of TF-IDF weight calculation, Linguistic Inquiry and Word Count (LIWC), part-of-speech approach~\cite{S2_Gachechiladze2017,S48_islam2019marvalous,S91_klunder2020identifying,S9_Cheriyan2021,S70_Sarker2020}.

\subsection{Sensor-based studies}
 23 out of the 64 primary studies utilise a variety of \textit{sensors} as the input of their machine learning models. The variety of sensors included \textit{skin conductance}, \textit{heart-related sensors}, \textit{muscle and nerve signals}, \textit{Oxygen rate}, \textit{respiratory signal}
 \textit{neural signals}. 

Most of these studies developed machine learning models with the emotional dimensions (including arousal and valence) as their dependent variables~\cite{S8_GirardiDaniela2020,S42_Nogueira2013,S46_GirardiLanubile2021,S79_fritz2016leveraging,S82_nogueira2015modelling,S86_koldijk2016detecting}. Two studies concern on one or more emotion detection, such as sadness~\cite{S84_epp2011identifying}, excitement~\cite{S55_kolakowska2013emotion,S84_epp2011identifying}, surprise~\cite{S55_kolakowska2013emotion}, and positive/negative emotion~\cite{S72_yang2021behavioral,S38_Muller2015}. Meanwhile, some studies also investigated on stress detection (Please see Table~\ref{tbl:_types_of_input}).

\textbf{Feature Extraction} -- Some studies used specific algorithms, techniques, or toolboxes to extract data from sensors. The algorithms utilsied includes cvEDA algorithm~\cite{S8_GirardiDaniela2020,S14_novielli2022sensor}, a band-pass algorithm~\cite{S8_GirardiDaniela2020,S14_novielli2022sensor}, Butterworth filter~\cite{S38_Muller2015} and a Weka tool~\cite{S38_Muller2015,S86_koldijk2016detecting}, Objective Player Experience Modelling (OPEM)~\cite{S42_Nogueira2013}, attention-based LSTM~\cite{S72_yang2021behavioral}, python package heart rate variability (HRV)~\cite{S78_rissler2020or}. Among all the techniques used in these studies, there were no common approaches utilised by different studies.

\subsection{Movement}
13 out of 64 papers employed movement-based input as the independent variables (Please see Table~\ref{tbl:_types_of_input}). However, all of these studies combined the variables with other variables such as text, sensor, utterances, and facial expression. These studies detected negative emotions (e.g., surprise, frustration, anger or sadness), positive emotions (e.g., happiness, excitement, etc.), and stress.

\textbf{Feature Extraction} -- There is no specific feature extraction used in this sample of studies. However, the variety of movement-based input physical activities (e.g. intensity of motion, energy expenditure, step counter)~\cite{S12_soto2021observing,S74_booth2022toward}, computer interaction (e.g., mouse and keyboard dynamics)~\cite{S44_androutsou2023automated,S55_kolakowska2013emotion,S75_padha2022quantum,S77_vrzakova2020affect,S81_anany2019influence,S83_naegelin2023interpretable,S84_epp2011identifying,S87_carneiro2012multimodal,S88_pepa2020stress,S89_alberdi2018using}.

All of these studies implemented different approaches to extracting the features.

\subsection{Utterances and facial expressions}
Finally, only 10 out of 64 papers employed either "Utterances" or "Facial expressions" as the independent variables. Please see Table~\ref{tbl:_types_of_input}. Only two studies took one type of input (e.g., facial expression)~\cite{S23_manikandan2024stress,S27_ballesteros2024facial} to detect stress and emotions (e.g., anger, sadness, fear, surprise, and disgust). The remaining studies combined these two variables and other variables such as sensor-based and movement-based inputs with the varieties of detection, including finite emotions(e.g., anger, sadness, fear, happy/love, surprise, excitement), continuous emotion, and stress. 

\textbf{Feature Extraction} -- 
For voice data, features were extracted by a specific tool developed by MIT~\cite{S28_awan2023creating} and analysed the verbal data with Shapley additive explanations (SHAP). Basic statistics functions were used for calculation as utterance-level features. Meanwhile, for facial expression data, facial landmarks (e.g. eye movement, brow furrows, lip curvature) were monitored~\cite{S23_manikandan2024stress}. Facial analysis system~\cite{S28_awan2023creating}, dynamic shape and the 3D mask models~\cite{S55_kolakowska2013emotion} were used to analysing non-verbal cues. An Affectiva's API was used to capture facial expressions~\cite{S81_anany2019influence}.

\subsection*{\textbf{Summary of findings in RQ2}} 
The findings for RQ2 indicate that machine learning models utilize a variety of inputs (text, sensor data, movement, utterances, and facial expressions) with text-based inputs being the most prevalent, often used for detecting emotions, stress, and toxic relationships, while other inputs are employed less frequently for specialized purposes such as enhancing emotional and stress detection.

\section{Findings -  (RQ3)}
\label{sec:_findings_rq3_methods}

From the results of the previous research question, we have found that those studies implemented various independent variables to classify different types of input data into sentiment (emotion) polarity, human emotion, emotional dimensions, stress, attrition, and toxicity polarity. The approaches employed in that collection of studies is machine learning also present their performances in predicting the classes. 

Given the variety of data sources and models, we closely followed the data clustering approach of the work presented in Hall et. al.,\cite{hall2011systematic}. Below, we report on the ML performances grouped by study purpose: we separate the “Emotion and stress detection using text”  from “Emotion and stress detection using sensors and movement”, given the large dissimilarity in input capture.

\subsection{ML Performances for Emotion Detection  (using text)}
\begin{sidewaysfigure}[htbp!]
    \centering
    \includegraphics[scale=0.5]{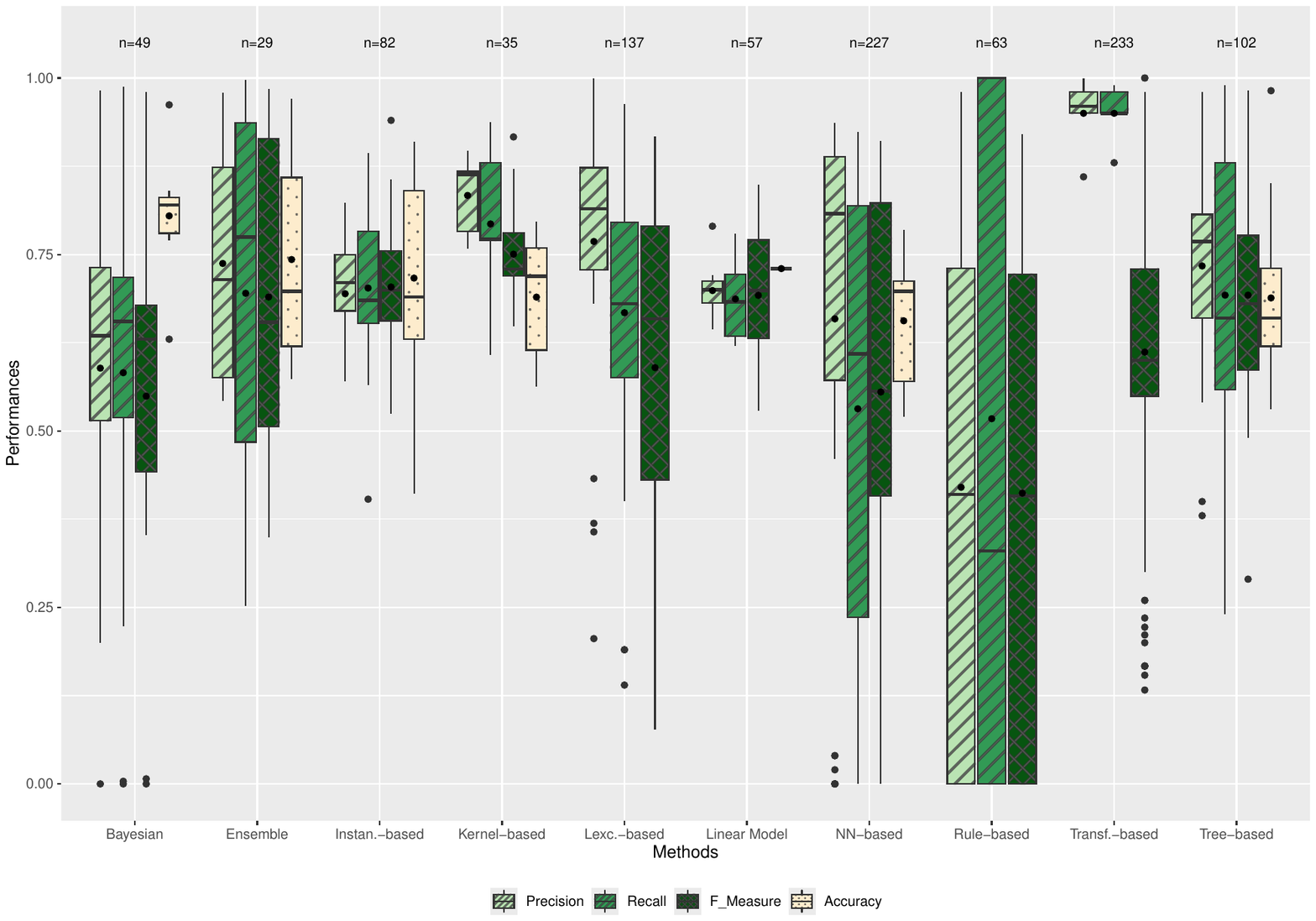}
    \caption{Model Performances of emotion and stress detection with machine learning techniques categorised based on their most generalised characteristics. Bayesian (e.g. Naïve Bayes), Lexicon based (e.g. Deva, Tensistrength, EmoTxt), Ensemble method (e.g. Adaptive Boast, XGBoost, CatBoost, Boosting), Instance-based algorithm (e.g. KNN), Linear Model (e.g. Logistic Regression), NN-based algorithms (e.g. MLP, SLP, Sentimoji, CNN), Ruled-based Algorithm (e.g. ZeroR), Tree-based Algorithms (e.g. DT, J48, RF), and Transformer-based Algorithms (e.g., GoEmotion, BERT, Albert, Roberta, Bert, Codebert, Graphcodebert)} 
    \label{fig:_emotion_detection_with_machine_learning}
\end{sidewaysfigure}

In this section, we report the performances of the machine learning models that aim to detect emotions using text as the main model input. Figure~\ref{fig:_emotion_detection_with_machine_learning} shows the boxplots representing the performances of precision, recall, f-measure, and accuracy of 10 different categories : Bayesian, Ensemble, Instance-based, Kernel-based, Lexicon-based, Linear Model, NN-based, Rule-based, Transformer-based, and Tree-based. We categorised the machine learning algorithms based on their most generalised characteristics. Bayesian include Naïve Bayes; Ensemble: Adaptive Boost, XGBoost, CatBoost, Boosting; Instance-based: KNN; Kernel-based: SVM; Lexicon-based: DEVA, Tensistrength, EmoTxt; NN-based: MLOP, SLP, SentiMoji, CNN; Ruled-based: ZeroR; Tree-based: DT, J48, RF, and Transformer-based: GoEmotion, BERT, Albert, Roberta, Codebert, Graphcodebert.

\textbf{Precision} - Concerning the precision of the performances in each machine learning method, the median values in all the methods range sparsely between 0.375 and 1. In detail, the Transformer-based method clearly shows the highest value, just above 0,9375. It is followed by Kernel-based, which is just below 0.875; and Lexicon-based, which is just below 0.8125. The Rule-based boxplot has the lowest value, below 0.5. Although the transformer-based group has a very small dispersion, it has asymmetrical skewness. Similarly, the Kernel-based set of experiments has wider dispersion with a very right-skewed distribution.

\textbf{Recall} - Concerning the recall of the performances, the median values range between 0.25 and 1 across all the techniques. In detail, the Transformer-based boxplot has the highest value at about 0.9375; this is followed by the Kernel-based, with a value above 0.75. Nevertheless, Rule-based has the lowest median value at about 0.3125. Out of 10 categories, four groups have a very narrow dispersion, including the Transformer-based, the Instance-based, the Kernel-based, and the Linear Model. However, the number of variants in the models, specifically the last three aforementioned models, is relatively low, with 82, 35, and 57 variants, respectively. Most of the techniques present asymmetric skewness, except the Lexicon-based. In general, in terms of recall measurement, Transformer-based models perform better than other methods. 

\textbf{F-measure} – Regarding the f-measure, the median values in all the methods range sparsely between 0.3125 and 0.75. In detail, 7 out of 10 groups has the median values above 0.625. These include Bayesian, Ensemble, Instance-based, Kernel-based, Lexicon-based, Linear Model, and Tree-based. Among these, Kernel-based has the highest value, above 0.6875. In terms of dispersion across all the methods, about 30\% of the approaches have narrow dispersion; these include Instance-based, Kernel-based, and Linear Model. Most of the batches in each category have asymmetrical skewness, with the exception that Instance-based, Linear Model, and Tree-based have otherwise. In general, regarding f-score measurement, Kernel-based methods perform better than other groups across the studies.

\textbf{Accuracy} – Not all the studies reported their accuracy performances. The bayesian group has the highest value, above 0.75; while the rest of the groups have values between 0.625 and 0.75. The dispersion of this measurement among all of the methods varies, with the Bayesian method having the narrowest spread. All of the categories has asymmetrical skewness. In general, in terms of accuracy measurement, the Bayesian method has the best accuracy compared to other methods.

\textbf{Number of variants} - Transformed-based category has the highest number of experiments, which is 233; this is followed by NN-based, Lexicon-based, and Tree-based, with 227, 137, and 102 experiments, respectively.

\begin{figure*}[htbp!]
    \centering
    \includegraphics[scale=0.6]{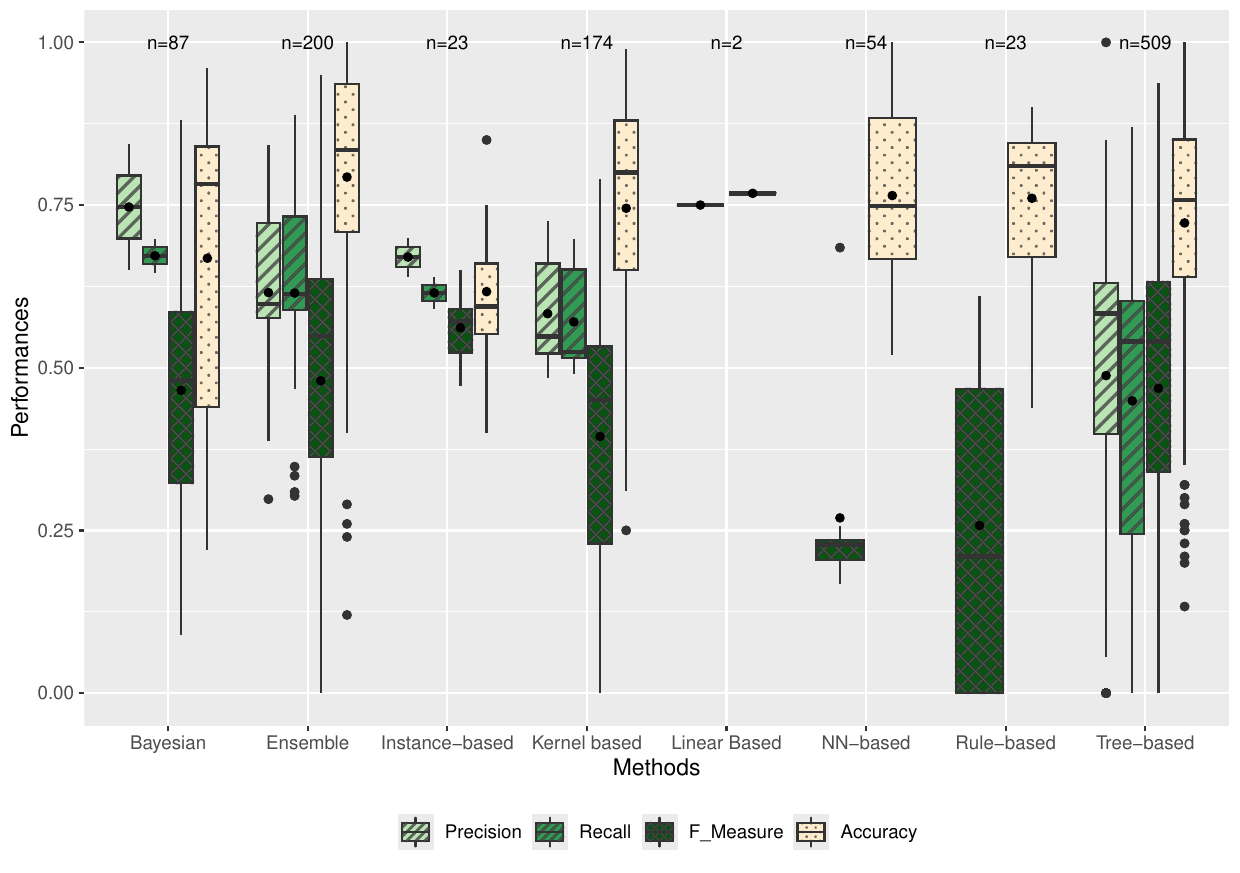}

    \caption{Model Performances of Emotion and Stress Detection with sensor and movement data as the inputs and machine learning techniques categorised based on their most generalised characteristics. Bayesian (e.g. Naïve Bayes, Bayes Net), Ensemble method (e.g., AdaBoast, Light Gradient Boasting Method), Instance-based algorithm (e.g. KNN, K-star, IBk), Linear Model (e.g., Logistic Regression), NN-based algorithms (e.g. CNN, RestNet), Ruled-based Algorithm (e.g. ZeroR), Tree-based Algorithms (e.g., DT, J48, RF, C.45)}
    \label{fig:_human_emotion_with_sensors}
\end{figure*}

\subsection{ML Performances for Emotion and Stress Detection  (using sensors and movement)}
\label{subsec:_Emotion_sensor_performances}
In this section, we report model performances of emotion and stress detection with sensor and movement data as the inputs and machine learning approaches, grouped by categories. Figure~\ref{fig:_human_emotion_with_sensors} shows the boxplots representing the performances of precision, recall, f-measure, and accuracy of 8 different categories: Bayesian, Ensemble, Instance-based, Kernel-basedLinear Model, NN-based, Ruled-based, and Tree-based. We categorised the machine learning algorithms based on their most generalised characteristics. Bayesian include Naïve Bayes, Bayes Net; Ensemble: Adaptive Boost, Light Gradien Boosting; Instance-based: KNN, K-star, IBk; Kernel-based: SVM; NN-based: CNN, RestNet; Ruled-based: ZeroR; and Tree-based: DT, J48, RF, C.45. One category, Linear Based(e.g. Logistic Regression), only shows two experiments (n=2); further NN-based and Rule-based categories show f-scores and accuracy.

\textbf{Precision} - Concerning the precision of the performances in each machine learning method, the median values in all the methods range sparsely between 0.5 and 0.75. In detail, Bayesian has the highest value, just below 0.75. This if followed by Instance-based, above 0.625. Meanwhile, Kernel-based has the lowest, at above 0.5. In terms of dispersion, Instance based has the narrowest distribution. Bayesian and Instance-based have symmetrical skewness. In general, Bayesian outperforms all the remaining methods.

\textbf{Recall} - The recall performances show median values ranging from 0.5 to 0.75 across all techniques. Bayesian lead with a median of above 0.625, followed by Ensemble and Instance-based models at just below 0.625, while Kernel-based models have the lowest median at just above 0.5. Among the 8 categories, two groups—Bayesian and Instance-based—exhibit very narrow dispersion and symmetrical spread. In summary, Bayesian models outperform other methods in recall measurements.

\textbf{F-measure} – Regarding the f-measure, most methods exhibit median values ranging between 0.125 and 0.625. Specifically, only one category, instance-based methods, has the highest median value. In terms of dispersion across all the methods, instance-based and NN-based have narrower distributions than the rest, even though the range of NN-based's dispersion falls between 0.125 and 0.25, which is significantly low. All of the batches in each method display asymmetrical skewness. In general, in terms of f-score measurement, Instance-based performs slightly better than other methods across the studies. 

\textbf{Accuracy} – Concerning the accuracy shown by the groups, the accuracy's median values fall between 0.5 and 0.825. Almost all the methods reported their accuracy above 0.75. Nevertheless, Instance-based has the lowest median value with its smallest dispersion. Ensemble has the highest accuracy with a wide dispersion. In general, in terms of accuracy measurement, Ensemble has the best accuracy compared to other methods. 

\textbf{Number of variants} - Although Tree-based group performances (e.g. precision, recall, and f-scores) show only in the range between 0.5 and 0.625, this group has the highest experiments, 509 in total, compared to the rest of the groups. This number of experiments is followed by the Ensemble group, with 200 experiments, and the Kernel-based, with 174 experiments.

\subsection{ML Performances for Attrition prediction}
\label{subsec:_Attrition_prediciton_performances}
\begin{figure}
    \includegraphics[scale=0.6]
    {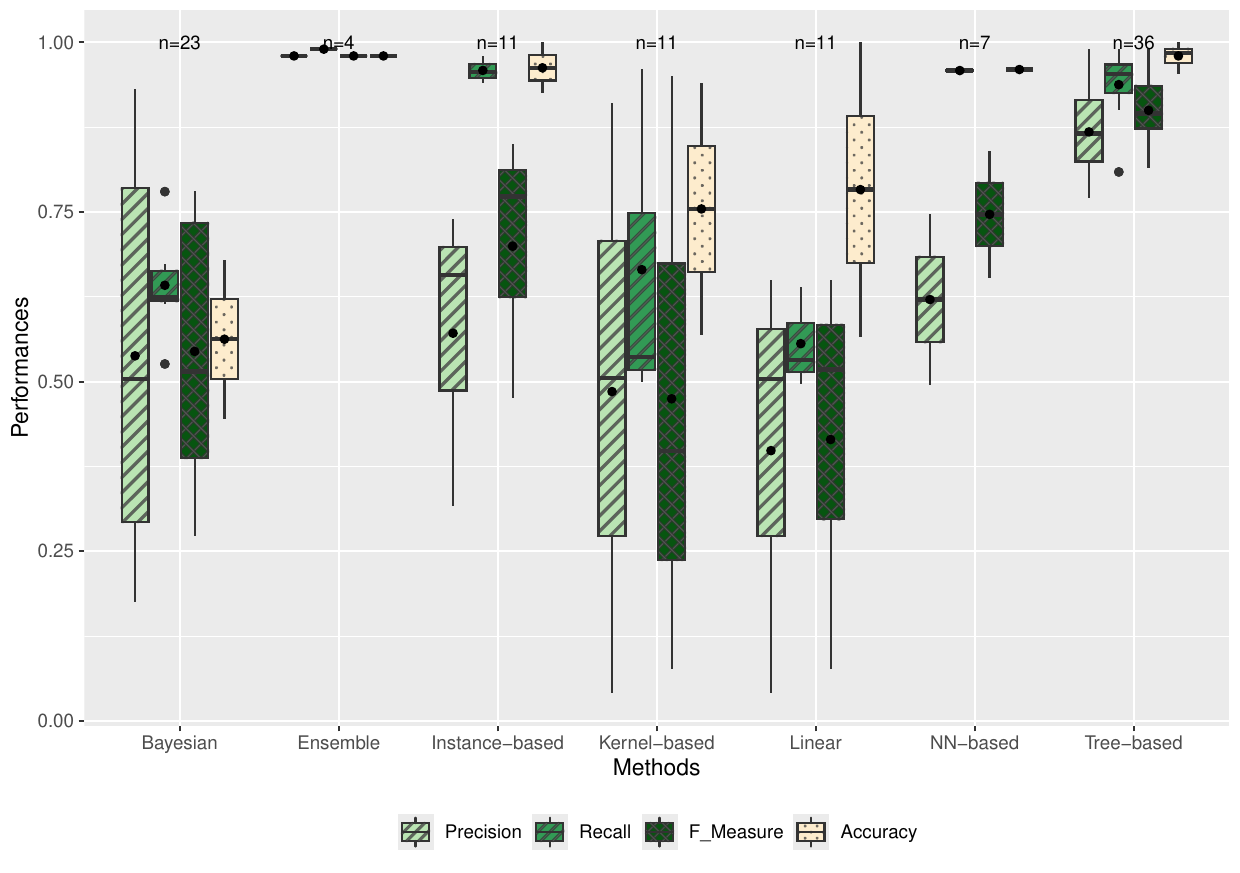}

    \caption{Model Performances of attrition detection with machine learning techniques categorised based on their most generalised characteristics. Bayesian (e.g. Naïve Bayes), Instance-based algorithm (e.g. KNN), Linear Model (e.g., Logistic Regression), NN-based algorithms (e.g. MLP), Tree-based Algorithms (e.g., DT, RF), Ensemble (e.g., Adaptive Boost), and Kernel-based algorithms (e.g., SVM, SVC)}
    \label{fig:_attrition_detection}
\end{figure}

In this section, we delve into the performances of various machine learning techniques used to predict attrition. Figure~\ref{fig:_attrition_detection} shows boxplots representing the performances of precision, recall, f-measure, and accuracy of 7 different categories: Bayesian, Ensemble, Instance-based, Kernel-based, Linear, NN-based and Tree-based. In details, Bayesian includes Naïve Bayes, Instance-based algorithms: KNN; Linear Model: Logistic Refression; NN-based algorithm: MLP; Tree-based algorithms: DT and RF; Ensemble method: Adaptive Boost; and Kernel-based algorithms: SVM, SVC.

\textbf{Precision} - Regarding the precision, the median values of all the groups vary, falling between just above 0.5 and 0.825. Tree-based group show the highest median value, just below 0.825. Moreover, this group also show a relatively narrow dispersion and symmetrical skewness. Similarly, NN-based own narrow and symmetrical skewness with the median value just below 0.625. Nevertheless, one experiment only with the Ensemble method shows a significantly high value, just below 1.
In general, Tree-based has the best performance in terms of precision. 

\textbf{Recall} – In terms of recall, the median values fall between 0.5 and 1. Four categories show high scores, above 0.825. These include Instance-based, Tree-based, NN-based and Ensemble-based, regardless number of experiments they have. TU put into detail, Instance-based and Tree-based have the narrowest dispersion, with their distribution are relatively normal. In general, these sets of experiments perform better in terms of recall.

\textbf{F-measure} – F-measure performances of all the categories range between 0.375 and 1. Tree-based and Ensemble have higher f-scores regardless number of experiments they have. Ensemble has the highest score, just below 1; nevertheless, the experiment has been conducted once. Meanwhile, Tree-based is a little bit lower than with the narrow dispersion and very symmetrical skewness. 

\textbf{Accuracy} – In terms of accuracy scores, the lower bound of the median values is higher, above 0.5, and the upper bound of the values is 1. In detail, Tree-based again shows the highest performance, with just below 1. This is followed by Ensemble with only one experiment, Instance-based, and NN-based with only one experiment. Tree-based and Instance-based have very narrow dispersion, but  Instance-based has a symmetrical skewness. Overall, Tree-based and Instance-based have the better performance compared to the rest of the boxplots. 

\textbf{Number of variants} -- This clearly shows that the number of experiments in each batch is relatively low, with 36 experiments the highest and 4 the lowest. Tree-based is employed more in the experiments, 36 experiments in total. This is followed by Bayesian which has 23 experiments in total. 

\subsection{ML Performances for Toxicity Detection}
\label{subsec:_toxic_performances}
\begin{figure}
    \includegraphics[scale=0.6]
    {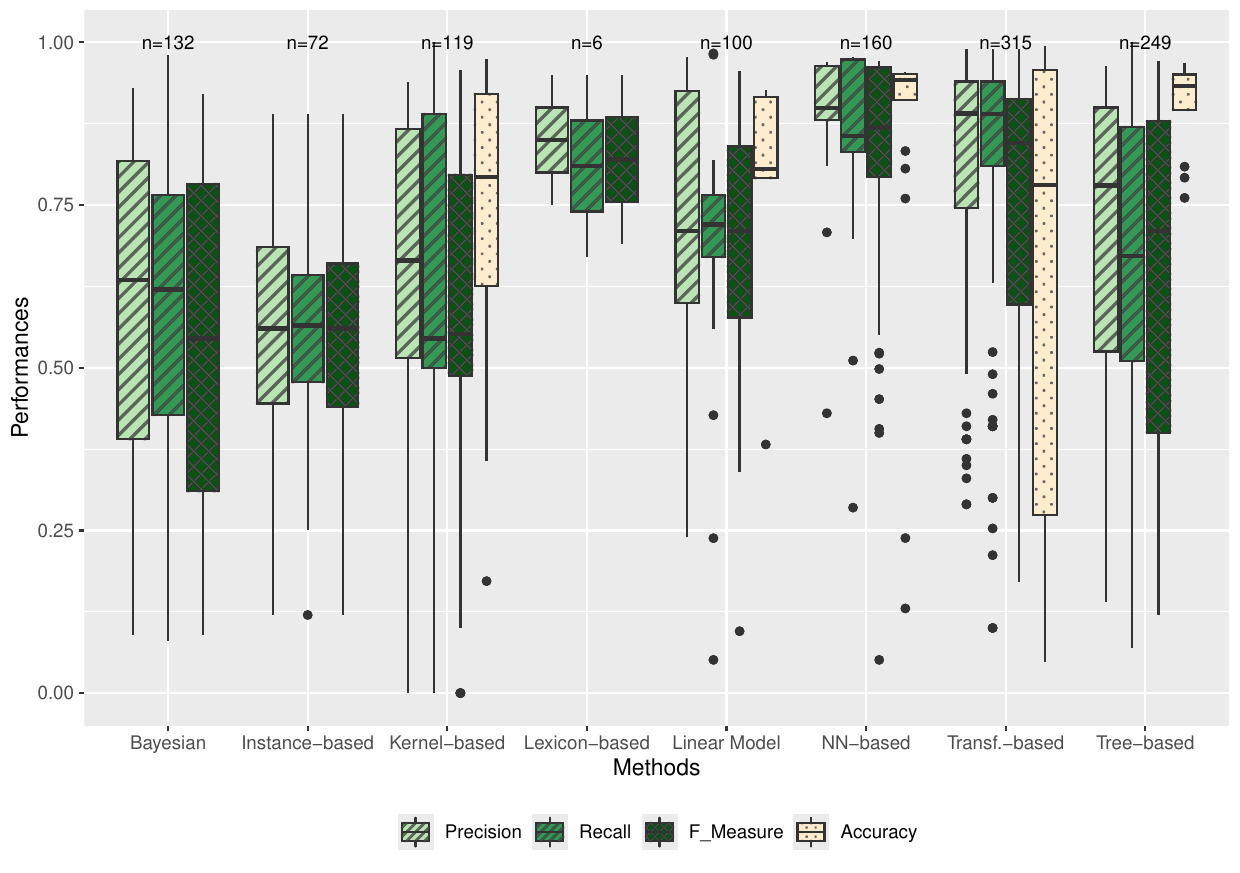}

    \caption{Model Performances of toxicity detection with machine learning techniques categorised based on their most generalised characteristics. Bayesian (e.g. Naïve Bayes), Instance-based algorithm (e.g. KNN), Linear Model (e.g., Logistic Regression), NN-based algorithms (e.g. Deep Pyramid CNN, Strudel, DPCNN, DCRNN, LSTM, BiLSTM, and GRU), Tree-based Algorithms (e.g., DT, GBT, RF, CART), and Transformed-based model (e.g., BERT, RoBERTa, DistilBERT, ALBERT, XLNet, ChatGPT, and GPT4.0)}
    \label{fig:_toxic_detection}
\end{figure}

In this section, we report model performances of toxicity detection. Figure~\ref{fig:_toxic_detection} shows the boxplots of 8 groups of methods consisting of Bayesian, Instance-based, Kernel-based, Lexicon-based, Linear Model, NN-based, Transformer-based, and Tree-based models. In details, Bayesian includes Naïve Bayes, Instance-based algorithm: KNN; Linear Model: Logistic Regression, NN-based algorithms: Deep Pyramid CNNN, Strudel, DPCNN, DCENN, LSTM, BiLSTM, and GRU; Tree-based algorithms: DT, GBT, RF, CART; and Transformer-based models: BERT, RoBERTa, DistilBERT, ALBERT, XLNet, ChatGPT, and GPT4.0.

\textbf{Precision} - The precision scores of median values fall within the range from 0.5 to 1. NN-based and Transformer-based models have the highest scores, above 0.825. Meanwhile, Lexicon-based and Tree-based perform lower than the aforementioned groups, between 0.75 and 0.825. In terms of their dispersions, Lexicon-based and NN-based have narrower dispersion than the rest of the boxplots. However, only Lexicon-based experiments spread normally. In general, NN-based methods show better performance, although a few outliers exist.

\textbf{Recall} - The recall scores are similar to the precision ones, from 0.5 to 1, with Transformer-based models show the highest values, just above 0.825. This is followed by NN-based and Lexicon-based. Although Transformer-based has narrow dispersion, Linear Model has the narrowest spread. Furthermore, in terms of skewness, three categories show normal distribution: Instance-based, Lexicon-based, and Linear Model. Although Transformer-based models seem to have high performance, the number of outliers of this batch is higher than the rest of the boxplots. 

\textbf{F-measure} – F-measure scores of median values fall between 0.5 and 0.825. Within the range from 0.75 to 0.825, three groups (e.g., Lexicon-based, NN-based, and Transformer-based algorithms) show their performances, in which NN-based show the highest values, just below 0.825. Nevertheless, among these aforementioned three groups, Lexicon-based has the narrowest dispersion and normal distribution, regardless the lowest number of experiments. In general, NN-based methods show better performance, regardless outliers they have.

\textbf{Accuracy} – Upon our observation, five groups report their accuracies. The models perform better in terms of their accuracy. This is evident that all the median values are higher than 0.75. It clearly shows that NN-based and Tree-based have the highest values, above 0.825. The dispersions of these models are relatively low, although the outliers persist and their data spread unevenly. 

\textbf{The number of variants} – It clearly shows that most of the categories have been conducted with over 100 experiments, in which the Transformer-based group shows its highest number of experiments (n=315). Tree-based and NN-based groups also show high number of experiments, 249 and 160, respectively. Only Lexicon-based models report significantly low experiments, 6 in total.

\subsection*{\textbf{Summary of findings in RQ3}} The findings from RQ3 reveal that machine learning techniques employed for detecting emotion, stress, attrition, and toxicity demonstrate varying levels of performance, with: i) text-based models (particularly Transformer-based) giving the best results in emotion detection; ii) Bayesian methods performing well in sensor-based emotion detection; iii) Tree-based models showing the best promises for attrition prediction, and iv) NN-based and Transformer-based methods achieving the highest accuracy in toxicity detection. Overall, we summarised all the performances of each group of machine learning performances in table~\ref{tbl:_performances_RQ2}.

\begin{table*}[htpb!]
\centering
\caption{Summary of performance measurements - RQ3}
\label{tbl:_performances_RQ2}
\footnotesize
\begin{tabular}{p{2.5cm}|p{2.5cm}p{1.8cm}p{2cm}p{1.8cm}}
 & Precision & Recall & F-score & Accuracy \\\hline
Emotion and Stress detection  (using text)   &   Transformer-based model, Kernel-based       & Transformer-based model, Kernel-based model      & Kernel-based  model   &   Bayesian       \\
\rowcolor{LightCyan} Emotion and Stress detection  (using sensors and movement) &   Bayesian
        & Bayesian      &  Instance-based model     &   Ensemble       \\
Attrition Prediction &  Tree-based model        & Tree-based model       &  Tree-based, Ensemble       &   Tree-based       \\
\rowcolor{LightCyan}Toxicity detection &  NN-based model, Transformer-based model         &  Transformer-based model      &  NN-based model       &  NN-based model, Tree-based model        \\\hline
\end{tabular}
\end{table*}

\section{Findings - (RQ4)}
\label{sec:_findings_rq5_datasets}
\begin{figure*}[htbp!]
    \centering
    \includegraphics[scale=0.6]
{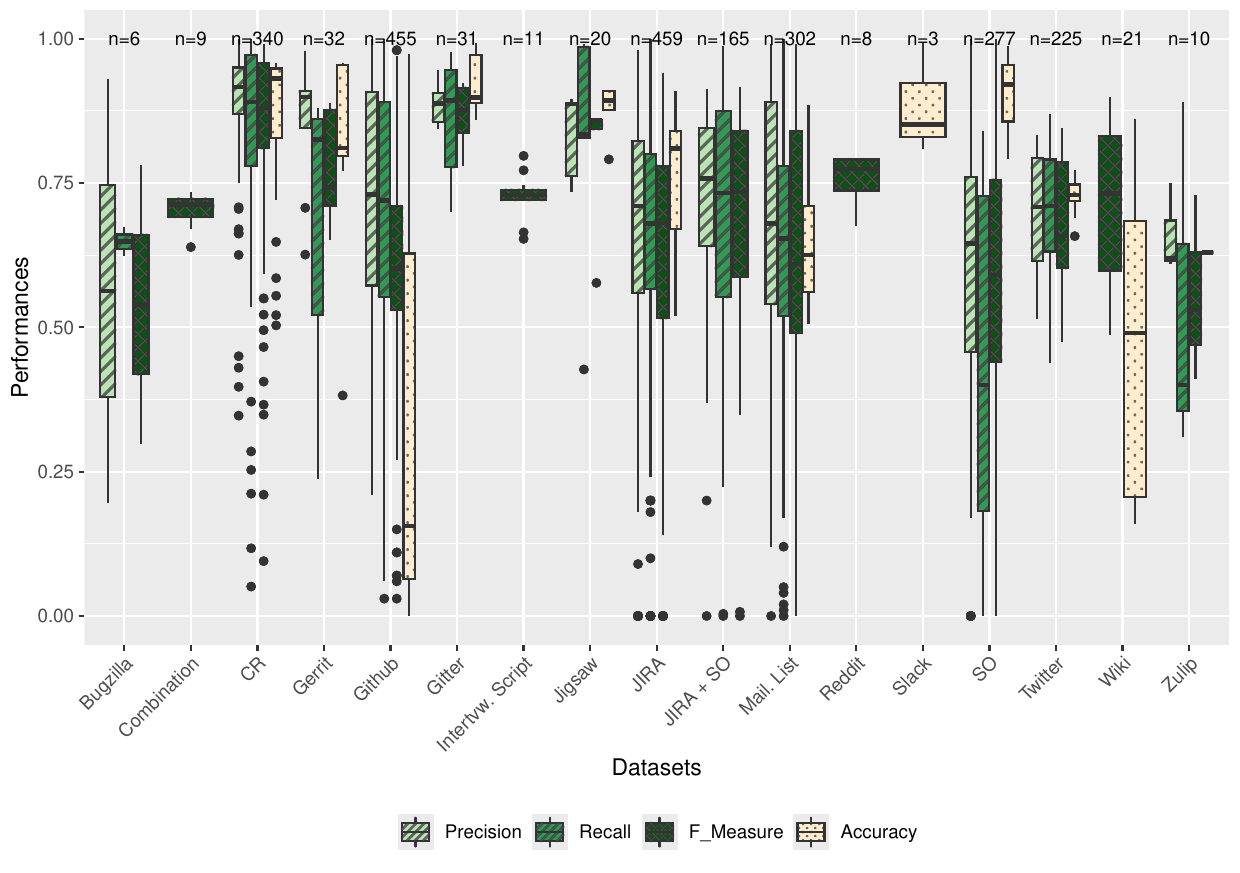}
    \caption{Model Performances by datasets}
    \label{fig:_datasets}
\end{figure*}

In this section, we summarise and visualise the performances of models using various types of datasets (see Figure~\ref{fig:_datasets}). We group the variants of models based on the datasets utilised in the studies, which include Bugzilla, Code Review (CR), Gerrit, Gitter, Interview Script, Jigsaw, JIRA, JIRA+SO, Mailing List, Reddit, Slack, Stack Overflow (SO), Twitter, Wiki, Zulip. We also added the "Combination" dataset, that is a group of mixed datasets consisting of DailyDialog, EmotionStimulus, and ISEAR dataset. ISEAR is a dataset consisting of sentences collected from a cross-cultural emotional response study across 37 countries. We report the performance metrics (e.g., precision, recall, F-measure, and accuracy) of each dataset. 

\textbf{Precision} - With regard to the performances, 11 out of 17 datasets have median values above 0.625  (e.g., CR, Gerrit, Github, Gitter, Jigsaw, JIRA, JIRA+SO, Mailing List, SO, and Twitter). Among these 11 datasets, Code Review, Gerrit, Gitter, and Jigsaw have their medians above 0.825. In addition, in terms of dispersion, CR, Gerrit, and Gitter show a narrow distribution compared to the other 8 boxplots.  Nevertheless, most of the datasets have an unbalanced distribution. In general, CR and Gitter show better performance, although the number of experiments with the Gitter dataset is only 31.

\textbf{Recall} - 
Concerning the recall of the performances, more than half of the datasets (10 out of 17) have a median value above 0.625 (e.g. Bugzilla, CR, Gerrit, Github, Gitter, Jigsaw, JIRA, JIRA+SO, Mailing List, and Twitter). Amongst these datasets, two of them, Code Review and Gitter, perform better than the rest ($> 0.825$). In terms of dispersion, Bugzilla has the narrowest and balanced distribution, regardless of the number of experiments done.

\textbf{F-measure} – 
Regarding the f-measure, the median values of most methods range between 0.625 and 1. In detail, two datasets: Code Review and Gitter have their median values above 0.825. In terms of dispersions across all the methods, Combination, Interview Script, Jigsaw, and Reddit have the narrow box, regardless of the low number of variants. None of the datasets has normal distribution, with the exception that Bugzilla show this normal spread. Overall, taking into account the number of variants along with their data distribution and median value, CR has the best performance among the rest.

\textbf{Accuracy} – Among all datasets that reported their accuracy, considering the median value and dispersion of the plot, Gitter, Jigsaw, and SO show good performance. In addition, SO, with its large number of experiments, has better data spread without any outliers than the other two datasets.

\textbf{The number of variants} – The plots show that the number of variants in each batch varies, with the lowest variants being 3 (e.g., Slack), and the highest being 459 experiments (JIRA). Although CR has a lower number of variants compared to JIRA, as well as a relatively high number of outliers, the median values of all performance metrics are above 0.825. 

\begin{table*}[htpb!]
\centering
\caption{Summary of performance measurements - RQ4. H, h, M, and L represent the performance value (median) with the criteria respectively: more than 0.8; between 0.7 and 0.8; between 0.5 and 0.7; below 0.5}
\label{tbl:_performances_RQ4}
\small
\centering
\begin{tabular}{l|llll}
                        & Precision & Recall & F-score & Accuracy \\\hline
Bugzilla             & M          &  M      &  M       &          \\
\rowcolor{LightCyan} Combination                  &           &        &         & M         \\
Code Review (CR)      &    H       &   H     &  H       & H         \\
\rowcolor{LightCyan}Gerrit            &  H         &  h      &  h       &   h       \\
Github                  &    h       &  h      &   M      &     L     \\
\rowcolor{LightCyan}Gitter                    & H          &  H      & H        &  H        \\
Interview Script &           &        &  h       &          \\
\rowcolor{LightCyan}Jigsaw            & H          & h       & h        &  H        \\
JIRA                   &       M    &     M   &     M    &      h    \\
\rowcolor{LightCyan}JIRA+SO          &   h        &   h     &   h      &          \\
Mailing List                   &   M        & M       &  M       &  M  \\
\rowcolor{LightCyan}Reddit          &           &        &   h      &          \\
Slack                   &           &        &         &   h \\
\rowcolor{LightCyan}SO          &   M        &   L     &   M      &   H       \\
Twitter              &   M        & M       &  M       &   h \\
\rowcolor{LightCyan}Wiki          &           &        &   M      &   L       \\
Zulip                   &   M        & L       &  M       &   M \\
\hline     
\end{tabular}
\end{table*}

\subsection*{\textbf{Summary of findings in RQ4}} Three datasets were found as the most promising for developing machine learning models, particularly those involving written communication: Code Review, JIRA and Gitter. Nevertheless, Gitter has been utilised in only a limited number of experiments. Some studies took into account mixed datasets such as JIRA+SO and Combination. The performance of the models using combination dataset is reported relatively good, between 0.625 and 0.75. Similarly, the performance of models utilising JIRA+SO is relatively good in terms of precision, recall, f-measure, and accuracy. Two studies that applied these combinations are done in~\cite{S34_maheshwarkar2021analysis,S48_islam2019marvalous}. We have summarized the performance measurements in Table~\ref{tbl:_performances_RQ4}.

\section{Discussion and Implications}
\label{sec:_discussions_and_implications}
\subsection{The trend of classification employed during the analysed period}
\begin{figure*}[htbp!]
    \centering
    \includegraphics[scale=0.6]{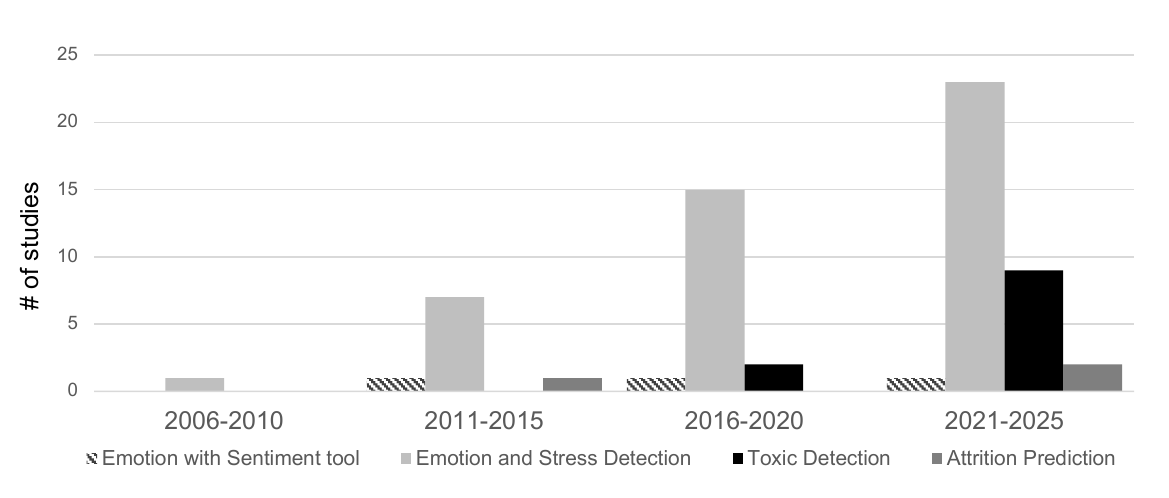}
    \caption{Trends of classification during the period}
    \label{fig:_trends}
\end{figure*}

According to RQ3 and RQ4, we analysed the frequency of studies performed within the three categories:  `Emotion and Stress Detection', 'Attrition Prediciton' and `Toxicity Detection', and during the period covered by the primary studies  (2001 - 2025). While the study presents insights between the year 2001 and 2025, the first evidence, relating to emotion detection with machine learning tools were reported in the year 2009 as was identified using the search. In Figure~\ref{fig:_trends} we plotted the 64 papers that employed machine learning in our primary studies, as clustered in the three categories. We divided the period covered by brackets of 5 years each: 2006 to 2010, 2011 to 2015, 2016 to 2020, and 2021 to 2025.

\textit{Emotion and stress detection} with machine learning  (with various types of input variables or independent variables) was the first type of study in its category, and it began to be pursued since 2009. This topic is clearly more popular than the other two types of detection, and as visible from the charts, it received increasing attention and multiple studies since its inception. Initially,  this approach utilised keystroke features (e.g., pause rate, timer per keystroke) through keyboard interactions, specifically to detect physical stress~\cite{S93_vizer2009automated}. In addition, this study also utilised linguistic features (e.g. lexical diversity, language complexity) to detect cognitive stress. Later in 2011, still using keystroke dynamics, a study done in~\cite{S84_epp2011identifying} to classify emotional states including nervousness, sadness, and tiredness. Only in 2013 a study by Nogueira et al.,~\cite{S42_Nogueira2013} extended the emotions to other classes: valence and arousal.

As an input to detect finite emotions such as anger and sadness, the text from the comments of agile-based projects started to be considered only in 2016~\cite{S40_ortu2016arsonists}. The popularity of written-based input continued until 2021~\cite{S4_murgia2018exploratory,S48_islam2019marvalous,S91_klunder2020identifying}. These studies collected various archived text: developer comments or reviews from JIRA and Stack Overflow, comments from the Apache Project, and chat messages (e.g., Zulip, Slack).

On the other hand, VAD detection  (as a particular case of emotion detection) started to be introduced in 2013. Nogueira et al.,~\cite{S42_Nogueira2013} utilised sensor data  (e.g psychophysiological metrics) as the input variables. All studies exploring VAD detection applied sensors as the inputs of their machine-learning models between 2015 and 2020~\cite{S8_GirardiDaniela2020,S77_vrzakova2020affect,S78_rissler2020or}. Independent variables used in the studies varied as we mentioned in section~\ref{sec:_findings_rq2_types_of_ml}. However, the study reported by M{\"a}ntyl{\"a} et al.,~\cite{S1_Mantyla2016} utilised issue reports of open source projects to recognise anger and sadness. This study calculated VAD scores to identify the emotions.

From the analysis of the primary studies, we also found that \textit{emotion/sentiment polarity} classification (“Emotion Detection with SA tools” in Table~\ref{fig:_trends}) began in 2014. A study conducted by Tourani et al.,~\cite{S51_tourani2014monitoring} analysed open-source mailing lists of Apache projects to investigate the sentiment among developers. This polarity classification was also used in detecting developers' emotions in 2016~\cite{S40_ortu2016arsonists}. This type of classification has been pursued continuously since, and the underlying study employed software developer's social media (e.g. tweets)~\cite{S31_silva2023using} as their datasets. 

The third category (“Toxicity Detection” in Figure~\ref{fig:_trends}) was first observed relatively recently: instead of classifying text into emotion polarity, the study conducted by Raman et al.,~\cite{S7_Raman2020} started to detect stress and burnout risk among the open-source developers by analysing Github issue comments in 2020. The study classified the issue comments into bipolar classes: toxic or non-toxic class. This study developed its proposed machine-learning models and considered different kinds of variables including politeness, subjectivity, sentiment, anger and comment length as their input variables. Similarly, a benchmark study was conducted in the same year to develop a large SE domain dataset used for toxic language detection~\cite{S70_Sarker2020}. In addition, studies on toxic relationships are getting more attention in recent year~\cite{S47_sarker2022identification,S16_Sarker2023,S60_mishra2024exploring,S73_rahman2024words}. This study evaluated some state-of-the-art toxic detectors on the proposed SE dataset. Detecting offensive language was also done by Cheriyan et al.,~\cite{S9_Cheriyan2021}. Using comments from Github, Gitter, Stack Overflow and Slack, this study manually classified the toxic comments into three classes (i.e., `personal', `racial', and `swearing') by employing a sentiment analysis approach, calculating Perspective API (PAPI) score, and obtaining Regular Expression (Regex) status.

The last category ("Attrition prediction") was first observed in 2013. A study done in~\cite{S37_Garcia2013} analysed the relationship between emotions and activity among OSS developers. Using Bugzilla and mailing list as its datasets, the authors built a bayesian-based classifier to predict the inactivity of developers. After a decade, few studies attempted to prevent burnout by predicting the attrition of employees. Using features extracted from surveys and available open-source datasets, these studies built machine-learning-based classifiers to predict employee attrition.

As we can see from the graph ~\ref{fig:_trends}, in the last five years, three types of classifications (`emotion detection with sentiment tool', `emotion and stress detection', `toxicity detection', and `attrition prediction') showed that researchers are working on different approaches to earlier recognise the risk of burnout among software developers. Those studies also showed that several types of features were extracted from the psychological and physiological domains, in order to i) enrich the variety of features used in the machine learning models and ii) to recognise the developers' behaviours. 

\subsection{Comparisons of machine learning-based models, methods, and datasets}
\subsubsection{Detecting emotion, stress, and attrition and toxic relationships with machine learning approaches}


The combination of Transformer-based models with online datasets (e.g. JIRA and Stack Overflow) to build a multi-class emotion classifier provides a practical suggestion to effectively detect early signs of burnout, in particular in detecting negative emotions, including sadness, anger, stress and depression. 

Furthermore, emotion detection models using sensors and movement as the input variables gave a good performance while the models were built with Bayesian, Instance-based, and Ensemble~(see table~\ref{tbl:_performances_RQ2}). However, the F-measure performances of these methods' models, as proposed in the previous studies, show unsatisfactory results: all of them score below 0.625~(see Section~\ref{subsec:_Emotion_sensor_performances}). The reason behind this is because detecting emotion (e.g. valence or arousal) is very different from classifying discrete emotion~\cite{ismail2023systematic}. The input variables, which are in continuous format, are completely different in terms of their dimensions. For example, heart rate and oxygen saturation have different ranges, which without proper normalisation or scaling may impact the F-score~\cite{ahsan2021effect,de2023choice}. In addition, the noise data contained physiological data may lead to more false positives and false negatives, ultimately lowering precision, recall, and F-Score~\cite{reiss2012introducing}.

The accuracies of the model's performances, which are above 0.75, in the studies provide encouraging signs of what method may be used to build the classifiers of toxic communication, and in particular, using a text-based input as the dataset. Although the Transformer-based models performed worse than their counterparts, the models mentioned in Section~\ref{subsec:_toxic_performances} may help in understanding the meaning of a word in the context of the entire sentence~\cite{vaswani2017attention}, leading to better contextual representations~\cite{S16_Sarker2023}. The models could also be pre-trained on large corpora of data, learning language representations such as text-based discussion during the software development cycle. For instance, the models trained in the large text-based discussion platforms, such as Github discussions, issues, or pull requests or other large platforms, which represent real-world cases, may improve the accuracy of the transformer models in real-world settings~\cite{jimenez2023swe,doan2023too}. 

Based on our findings, we advocate for the use of Decision Tree and Random Forest as the optimal algorithms for predicting attrition. The performance metrics illustrated in Figure~\ref{fig:_attrition_detection} demonstrate that these methods surpass the other techniques employed in the primary studies~\cite{jacobsen1999comparison,montgomery2024comparative,verma2025explanation}. This advantage may stem from the datasets utilised in those studies, which were obtained via surveys and relied on tabular data as input~\cite{ali2012random}. Nevertheless, we speculate that the results may be different with high-dimensional or unstructured data, such as free-text responses. Using our findings described in Figure~\ref{fig:_attrition_detection}, we believe the results may not be satisfactory.

The combination of the aforementioned methods with certain datasets, including Code Review, Gitter, Jigsaw, JIRA and Stack Overflow, is now a standard type of study, as these datasets are commonly used as the communication medium among the developers. In addition, Slack, Twitter and other communication media can also be taken as a data source in detecting toxic relationships and attrition prediction. These are widespread tools and provide a more natural way of communication among the developers. 


\subsubsection{Metrics considerations for evaluating model performances}
Some studies in emotion and stress detection reported accuracy metrics. These studies, which utilised sensor and movement data as input, favored accuracy as a straightforward metric representing the percentage of correct predictions. This choice likely reflects a preference for simplicity, as accuracy can effectively communicate model performance to non-technical stakeholders. Additionally, using accuracy is reasonable when the same subjects are present in both the training and test datasets~\cite{juba2019precision,S8_GirardiDaniela2020}. 

Since the accuracy metric is not ideal for imbalanced datasets, other metrics, such as the F1-score (which combines precision and recall) are recommended. In text-based datasets used reported in our findings, these metrics were particularly suited due to the nature of text-based datasets. Additionally, if evaluating a model’s predictive performance by class is of interest, the F1 score is preferable. However, if false negatives are considered more critical than false positives, recall should be prioritised over precision~\cite{yu2018total,grossman2016trec}. Conversely, if false positives carry greater weight, precision is recommended~\cite{zhang2007comments,gray2011further}. It is worth noting, however, that the F1 score may be less suitable when working with balanced datasets.

In conclusion, we recommend using accuracy metrics alongside precision, recall, and F-measure to provide a comprehensive review of machine learning models. Relying solely on one metric is insufficient, as each offers unique insights. For instance, our findings on toxicity detection models reveal that while many accuracy scores exceed the median threshold of 0.75, the medians of the other metrics fall below this level. This suggests that, although the models generally perform well, they are affected by imbalanced datasets. Therefore, it is crucial to measure precision and recall further, as these metrics highlight areas where tuning could enhance model performance~\cite{zhang2007comments,yu2018total,grossman2016trec,cormack2016scalability}.

\subsection{Considering other independent variables as predictors}
We further examined the studies that employed machine learning methods, focusing on other independent variables such as the number of participants, gender, and age. We found that only a few studies reported gender and/or age information in their experiments~\cite{S42_Nogueira2013,S8_GirardiDaniela2020,S12_soto2021observing,S93_vizer2009automated}, with participant numbers ranging from 10 to over 100, and a higher number of male participants compared to females. Among these studies, several also reported additional details such as years of experience, type of profession, and specific job functions~\cite{S42_Nogueira2013,S12_soto2021observing,S8_GirardiDaniela2020}. However, this demographic information was not used as input data for the machine learning models, and none of the studies presented results related to these demographic factors.

While our findings show that most performance metrics (e.g., F-measure and accuracy) exceed 0.5 (see Figure~\ref{fig:_human_emotion_with_sensors}), adding demographic factors as predictors in machine learning models may produce valuable insights. For instance, incorporating age could support more personalised models, as research indicates that older adults experience fewer and less intense stressors, along with a lower level of negative affect compared to younger adults~\cite{brose2015older,koffer2016stressor}. Additionally, including gender as a factor may enhance model personalisation, given findings that females tend to express emotions more openly than males~\cite{chaplin2015gender,deng2016gender}. Previous research in other fields suggests that gender and age could influence the likelihood of experiencing symptoms of burnout~\cite{marchand2018age,purvanova2010gender,jalili2021burnout}. However, these factors might also introduce bias or lead to only marginal improvements in performance.  It is also important to be mindful of ethical considerations, especially when using sensitive attributes like gender.

\subsection{Symptoms of exhaustion, cognitive dysfunction, and lack of pleasure in work}
Our findings suggest that certain factors could aid in the early identification of burnout symptoms. For instance, finite emotions (such as anger, anxiety, or frustration) can be detected with machine learning models before they escalate. Frequent bouts of frustration at work can accumulate and lead to emotional exhaustion, which is strongly associated with experiencing negative emotions. Individuals often report feeling exhausted, overwhelmed, and emotionally drained due to burnout~\cite{alessandri2018job,holmqvist2006burnout,liu2021negative,schoeps2019effects}.

Additionally, continuous emotions, particularly higher arousal levels seen in individuals with burnout and depression, may signal cognitive overload or dysfunction, impairing their ability to manage stress effectively~\cite{Haug2022}. This finding is supported by Kim et al., who reported that high arousal is likely linked to impaired stress responses and cognitive overload~\cite{kim2022technostress}.

Our study also emphasises the importance of internal human factors, such as emotional dimension states (e.g., low valence and high arousal), which can be identified using machine learning models. Detecting these emotional states may indicate reduced job satisfaction and engagement, leading to a lack of enjoyment in work tasks. This is consistent with other research~\cite{P72kaur2022didn,masri2023mental}, which highlights that valence and high arousal are significant predictors for monitoring employees' emotional states and predicting burnout risk.

On one hand, it is also evident that the prolonged stress over a certain time may lead to the risk of experiencing burnout~\cite{maslach2006burnout,maslach2001job}. To prevent this risk, detecting the stress symptoms earlier may be one of the recommendations to mitigate the burnout risk, as stress may have a positive relationship with burnout~\cite{bruce2009recognizing,maslach1997maslach,P36sonnentag1994stressor}. On the other hand, employee attrition may be a consequence of an individual on the verge of burnout~\cite{o2008burnout,chakrabarti2022more,stehman2019burnout}. Hence, predicting employee attrition may also be recommended to mitigate the risk of burnout. 

\begin{figure*}[htbp!]
    \centering
    \includegraphics[scale=0.8]{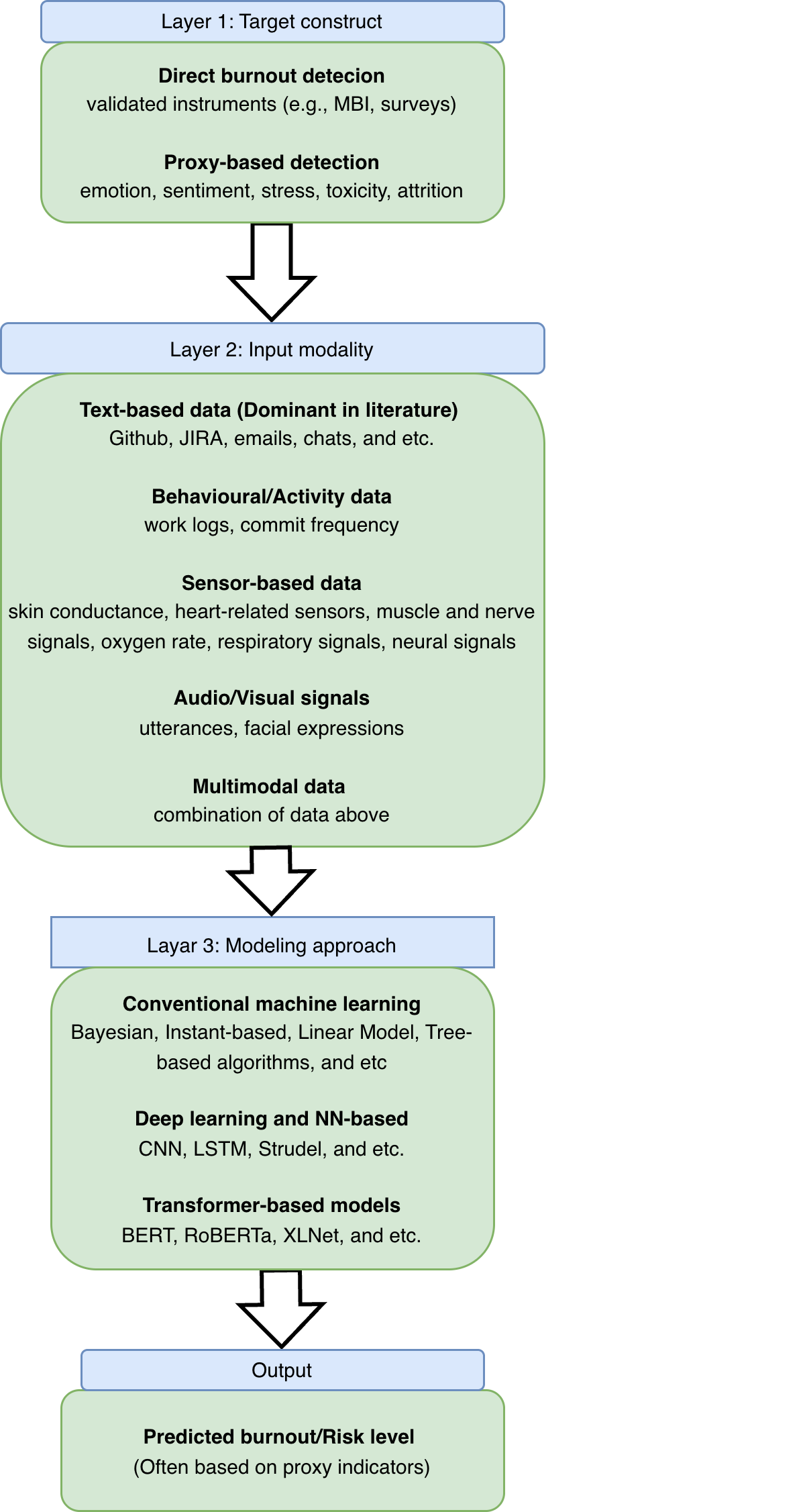}
    \caption{A framework of machine-learning based early detection of burnout}
    \label{fig:_framework_detection}
\end{figure*}

It is crucial to recognise certain possible benefits of putting machine learning tools into practice. Our literature reviews clearly show that the ML-based tools or models may give hints that people show early signs of burnout such as negative emotion~(anger), stress, frustration, and toxic relationship ~(see table~\ref{tbl:_table_synthesis}). Having a system that predicts early signs of burnout can contribute to taking measures that prevent burnout in some if not all.  As such, it can help to mitigate the burnout rates. Furthermore, early detection from the models may notify the developers or the managers to take precautionary actions such as taking temporary breaks and organising workload. As an indirect result, this may increase the productivity of the developers~\cite{kaur2020optimizing}. 

\subsection{A framework for machine learning-based burnout detection}
We propose a framework organizes the literature on machine learning-based burnout detection into three layers, namely: (1) the target construct being modeled, (2) the type of input data employed, and (2) the modelling techniques applied. The framework depicted in Figure~\ref{fig:_framework_detection} highlights a dominant reliance on proxy-based constructs and text-based, revealing a gap between machine learning practices and validated burnout measurement.

\subsection{Bridging the gap between theoretical research and practical applications}
There are several ramifications when considering the practical implications of machine learning models in the early detection of burnout: below we analyse the ones that we deem the more important ones.

1) \textit{Early warning systems.} Machine learning approaches can be instrumented to examine diverse data sources, and to detect patterns that may signal potential burnout. These identified warning signs can serve as an early alert system, for instance notifying managers and employers when indications of burnout are present.  The system would alert managers to potential burnout cases early on, allowing for timely intervention, such as redistributing workload or providing support to those at risk. In addition, individuals may notice early signs of burnout and take preventive steps to mitigate the risk such as taking breaks, setting boundaries, and incorporating stress-reducing activities into their routine.

Machine learning models may be trained on various types of data to produce potential burnout patterns at different phases: for example, during software development, various means are used to ease the collaboration (e.g. Git, Confluence, Google Doc), communication (e.g. Maililing List, Slack, Discord and video conferencing tools), and tracking of the progress of the project by both individuals and teams (e.g. Code Review tools, Jira, Clockify). 

2) \textit{Enhanced team dynamics.} Unhealthy communication among developers may lead to collaboration issues and potential other stressors. Machine learning models may analyse this type of communication patterns during software development within teams. As a result, the patterns obtained by the models can offer insights into the symptoms of stress derived from the collaboration. In addition, ML-driven insights can notify team-building strategies to foster a positive and supportive work environment, which may attract new developers to join the project and feel welcome in the new environment~\cite{constantino2023perceptions}.

3) \textit{Optimizing Workload and Task Allocation.} Analysing historical data may also be conducted by machine learning models to optimise workload distribution~\cite{shenouda2023improving,samir2023improving}, preventing excessive demands on certain team members. For instance, looking at the developers' commit history may reveal how frequently they are working. More work can mean more workload and this may lead to immense pressure on projects. Furthermore, machine learning models can offer insights to understand individuals and weaknesses, and the insights may recommend better task allocation, ensuring a more balanced and manageable workload.

One example of integrating an ML-based tool into an existing project management system is by developing APIs to connect the tool with the project management system. The APIs would facilitate the exchange of data and predictions between the two systems. On one side, the tool would collect various data points (e.g. work hours, communication patterns and logs, task completion rates, and any additional indicators of team member well-being) within the management system. The aim would be to identify patterns indicative of increased stress level or potential burnout. On the other side, the project management system would always be capable of providing the aforementioned data needed, as those are routinely collected by that system.

Nevertheless, several costs of implementing the models should be taken into account. The decision-makers may underestimate the initial investment in technology, software and possibly specialised expertise. For example, implementing machine learning models requires training on large volumes of unstructured and unlabeled data in various formats, such as text, images, or audio. Additionally, these models need to be optimised for speed to handle and process extensive datasets efficiently. Proper storage and deployment are also crucial, which often involves using cloud-based services or on-premise servers, both of which can be expensive. Moreover, it's important to consider the costs associated with hiring specialists—such as data scientists, data engineers, psychologists, or HR experts—to develop, maintain, improve, interpret, and refine the models~\cite{napoli2023cost}.

In addition, monitoring one developer's activity and maintaining the models themselves as a long-term investment should also be calculated~\cite{McKinsey2023}. As previously mentioned, potential costs for infrastructure and personnel are significant, but ongoing expenses for model maintenance must also be considered. For example, there may be costs associated with subscriptions to licensed tools used for logging, error tracking, and alerting within a machine learning system.

\textbf{Integration challenges and strategies} - On top of the aforementioned implications, it is worth mentioning a number of challenges and strategies in implementing machine learning models as tools to detect early symptoms of burnout. Combining data from various sources can be complex due to their differences in format and structure. The strategy to address these differences would be to develop standardised data formats and invest in tools that facilitate seamless data integration. This solution would also provide privacy-preserving techniques such as privacy-preserving data mining~(PPDM), secure informed consent~\cite{mendes2017privacy}, and comply with regulations regarding data protection.

The data privacy is one of the issues, given the concern that this important privacy should be echoed among the users beforehand. The clarity of agreement between the top leaders and the employees is key. Additionally, it is essential to ensure that the machine learning system complies with relevant laws and regulations, such as the GDPR for data privacy in the EU and HIPAA in the U.S. healthcare industry.

Finally, potential biases of the models and their results should also be considered as an important concern. The complexity of parameters in measuring burnout clearly shows that measurements cannot be considered from just one perspective. Additionally, early burnout signs prompted by the models may incorrectly predict the real emotions of the user; particularly, for professionals such as nurses, stewardesses, or customer service officers that involve more emotional labour~\cite{bani2021behind,colonnello2021reduced}. In short, machine learning models designed by data scientists must involve mental health and human resource professionals to create a multidisciplinary approach that considers both technical and human aspects.  

\section{Threats to validity}
\label{sec:_threats}
In this paper, we have made every effort to mitigate the rising challenges to the validity of a standard SLR. Below, we highlight the most significant ones, as they pertain to the phases of our SLR.

\textbf{Threats to construct validity -}
A key threat to construct validity in our search strategy arises from the potential mismatch between the search terms used and the underlying concepts we aim to capture. While our initial query focused on "automatic OR detection" within the context of software development and burnout, expanding the search to include additional terms like "stress," "emotion," and "sentiment" may introduce broader or tangential studies that do not directly address developer burnout or its automatic detection. This may return irrelevant or only loosely related papers. Hence, it may dilute the construct we intend to study. To mitigate this, we reevaluated the papers with our selection procedure described in our methodology. Additionally, some relevant studies may use different terminology not covered by our query, leading to their unintended exclusion. 

Regarding the online libraries used in our searching phase, we also utilised Google Scholar to broaden our search results: despite the number of false positives (noise), it has the ability to significantly increase the reach of the systematic search. Other databases, such as Scopus, may offer suboptimal results, according to reports~\cite{martin2018google}.

\textbf{Threats to internal validity} - The papers to be included in the set to be studied were chosen manually and with the aid of papers obtained automatically throughout the search process. To prevent bias, the first and third authors evaluated the articles to be included in the following round simultaneously, and discrepancies were resolved by discussion.

The literature listed by the snowballing method is very reliant on the initial start set used to identify the set and on a single query. Consequently, there may be additional important articles missing that can be identified with more precise and specialised database search phrases. 

The extensive analysis of the papers (title, abstract, and full text) was performed manually by the first author, who also reviewed each paper's complete text. Through discussion with the third author, inconsistencies or uncertainties were resolved.

\section{Conclusion and future works}
\label{sec:_conclusion}
We performed a systematic literature review of studies that proposed machine learning approaches. In particular, we focused on papers detecting burnout in software developers and IT professionals. 

Out of 64 relevant studies found in the literature search, we reviewed in depth this pool of papers that utilised machine learning methods, and with three main purposes: (i) detecting emotions and stress, (ii) detecting attrition, and (iii) detecting toxic relationships. 
We also identified four input types (e.g., text-based input, sensors-based input, utterances-based input, movement-based input and facial expressions) used in the ML models within the 64 studies.

Furthermore, we fully reviewed the accuracy, precision, recall, and F-score of the proposed ML methods for 64 studies. We reported the better ML-based modelling techniques, grouped by classification type (e.g. emotion polarity, stress detection, attrition detection and toxic detection). In addition, we reported the performance measurements by the datasets used in the studies.

Based on our findings, we identified the ML models that have been shown to perform better in the early detection of burnout. These may be combined with the off-the-shelf datasets in order to extend or develop machine-based classifiers, and in the context of detecting developers' emotions and toxic relationships. 


Future works will need to include big data, larger and more complex databases, as the input of machine learning models. 
This rich data, along with the emerging technology in artificial intelligence models, such as deep learning models and pre-trained large language models, will likely boost the performance of early-burnout detection models. 

Additionally, emerging trends such as the integration of multi-modal data using AI can offer a comprehensive view of a person's emotional state. Research could focus on combining data from various sources—such as text-based communication, biometric data (e.g., heart rate, skin conductance), and facial recognition—to enhance burnout prediction models.

Advancements in sensor technologies, including wearable devices and brain-computer interfaces (BCIs), enable more precise real-time monitoring of emotional and physiological states. Future research could investigate how continuous monitoring with these advanced sensors might provide earlier detection and enable personalised intervention strategies for burnout.

 Furthermore, the increased accessibility to social media will be increasingly considered as another type of source which may give insights into what is happening to software engineers in their daily life. Some studies investigated social media posts and revealed the likelihood that an individual suffered from depression or stress~\cite{ghosh2021depression,ahmed2022machine}. In term of software engineering domain, extending the data source to this media will enrich the variety of data used by machine learning models.

Finally, future work may delve into the exploration of predictive analytics and personalised AI models. These models would consider individual differences in stress responses and their unique working styles, aiming to not only detect but also predict the likelihood of burnout.

\section*{Declarations}
\subsection*{\textbf{Funding}}
This study was supported by the Indonesia Endowment Fund for Education (LPDP), Ministry of Finance of Republic of Indonesia, Ref. Number S-943/LPDP.4/2021.

\subsection*{\textbf{Ethical approval}}
Ethical approval: not applicable

\subsection*{\textbf{Informed consent}}
Informed consent: not applicable

\subsection*{\textbf{Author Contributions}}
\textbf{Tien Rahayu Tulili:} Conceptualization, Literature Search, Methodology, Data Analysis, Writing--Original Draft, Revision, Visualization. \textbf{Ayushi Rastogi:} Conceptualization and Writing--Review, Revision. \textbf{Andrea Capiluppi:} Conceptualization, Analysis, Writing--Review and Editing.

\subsection*{\textbf{Data Availability Statement}}
The data associated with this study are publicly available online in the replication package\footnote{\url{https://github.com/phd-work-22/SLR-Early-Identification-of-Burnout/tree/main}}.

\subsection*{\textbf{Conflict of Interest}}
The authors have no competing interests to declare that are relevant to the content of this article.

\subsection*{\textbf{Clinical Trial number}}
Clinical trial number: not applicable
\bibliographystyle{spmpsci}      
\bibliography{bibliography}   

@article{maslach2006burnout,
  title={Burnout},
  author={Maslach, Christina and Leiter, MP},
  journal={Stress and quality of working life: current perspectives in occupational health},
  volume={37},
  pages={42--49},
  year={2006},
  publisher={IAP}
}

@inproceedings{wohlin2014guidelines,
  author		= "Wohlin, Claes",
  title			= "Guidelines for snowballing in systematic literature studies and a replication in software engineering",
  editor		= "n/a",
  volume		= "n/a",
  booktitle		= "Proceedings of the 18th international conference on evaluation and assessment in software engineering",
  pages			= "1--10",
  address		= "Heidelberg",
  publisher		= "Springer",
  year			= "2014",
  doi = "10.1145/2601248.2601268"
}

@inproceedings{S1_Mantyla2016,
author = {M\"{a}ntyl\"{a}, Mika and Adams, Bram and Destefanis, Giuseppe and Graziotin, Daniel and Ortu, Marco},
title = {Mining Valence, Arousal, and Dominance: Possibilities for Detecting Burnout and Productivity?},
year = {2016},
isbn = {9781450341868},
publisher = {Association for Computing Machinery},
address = {New York, NY, USA},
doi = {10.1145/2901739.2901752},
booktitle = {Proc. of 13th Intl. Conf. on Mining Software Repositories},
pages = {247–258},
numpages = {12},
location = {Austin, Texas},
series = {MSR '16}
}

@inproceedings{S2_Gachechiladze2017,  
author = {Gachechiladze, Daviti and Lanubile, Filippo and Novielli, Nicole and Serebrenik, Alexander},  
booktitle = {2017 IEEE/ACM 39th International Conference on Software Engineering: New Ideas and Emerging Technologies Results Track (ICSE-NIER)},   
title = {Anger and Its Direction in Collaborative Software Development},   
year={2017},  
volume={},  
number={},  
pages={11-14},  
doi={10.1109/ICSE-NIER.2017.18},
publisher = {IEEE/ACM},
address = {USA}
}

@article{S4_murgia2018exploratory,
  title={An exploratory qualitative and quantitative analysis of emotions in issue report comments of open source systems},
  author={Murgia, Alessandro and Ortu, Marco and Tourani, Parastou and Adams, Bram and Demeyer, Serge},
  journal={Empirical Software Engineering},
  volume={23},
  number={1},
  pages={521--564},
  year={2018},
  publisher={Springer},
  doi = {10.1007/s10664-017-9526-0}
}

@inproceedings{S6_Islam2018,
author = {Islam, Md Rakibul and Zibran, Minhaz F.},
title = {DEVA: Sensing Emotions in the Valence Arousal Space in Software Engineering Text},
year = {2018},
isbn = {9781450351911},
publisher = {Association for Computing Machinery},
address = {New York, NY, USA},
doi = {10.1145/3167132.3167296},
booktitle = {Proc. of the 33rd ACM Symposium on Applied Computing},
pages = {1536–1543},
numpages = {8},
location = {Pau, France},
series = {SAC '18}
}

@inproceedings{S7_Raman2020,
author = {Raman, Naveen and Cao, Minxuan and Tsvetkov, Yulia and K\"{a}stner, Christian and Vasilescu, Bogdan},
title = {Stress and Burnout in Open Source: Toward Finding, Understanding, and Mitigating Unhealthy Interactions},
year = {2020},
isbn = {9781450371261},
publisher = {Association for Computing Machinery},
address = {New York, NY, USA},
doi = {10.1145/3377816.3381732},
booktitle = {Proceedings of the ACM/IEEE 42nd International Conference on Software Engineering: New Ideas and Emerging Results},
pages = {57–60},
numpages = {4},
location = {Seoul, South Korea},
series = {ICSE-NIER '20}
}

@inproceedings{S8_GirardiDaniela2020,
author = {Girardi, Daniela and Novielli, Nicole and Fucci, Davide and Lanubile, Filippo},
title = {Recognizing Developers' Emotions While Programming},
year = {2020},
isbn = {9781450371216},
publisher = {Association for Computing Machinery},
address = {New York, NY, USA},
url = {https://doi.org/10.1145/3377811.3380374},
doi = {10.1145/3377811.3380374},
booktitle = {Proceedings of the ACM/IEEE 42nd International Conference on Software Engineering},
pages = {666–677},
numpages = {12},
location = {Seoul, South Korea},
series = {ICSE '20}
}

@incollection{S9_Cheriyan2021,
author = {Cheriyan, Jithin and Savarimuthu, Bastin Tony Roy and Cranefield, Stephen},
title = {Towards Offensive Language Detection and Reduction in Four Software Engineering Communities},
year = {2021},
isbn = {9781450390538},
publisher = {Association for Computing Machinery},
address = {New York, NY, USA},
url = {https://doi.org/10.1145/3463274.3463805},
doi = {10.1145/3463274.3463805},
booktitle = {Evaluation and Assessment in Software Engineering},
pages = {254–259},
numpages = {6},
location = {Trondheim, Norway},
series = {EASE 2021}
}

@article{S12_soto2021observing,
  title={Observing and predicting knowledge worker stress, focus and awakeness in the wild},
  author={Soto, Mauricio and Satterfield, Chris and Fritz, Thomas and Murphy, Gail C and Shepherd, David C and Kraft, Nicholas},
  journal={International Journal of Human-Computer Studies},
  volume={146},
  pages={102560},
  year={2021},
  publisher={Elsevier},
  doi = {10.1016/j.ijhcs.2020.102560}
}

@conference{S13_qiu2022detecting,
  title={Detecting Interpersonal Conflict in Issues and Code Review: Cross Pollinating Open-and Closed-Source Approaches},
  author={Qiu, Huilian Sophie and Vasilescu, Bogdan and K{\"a}stner, Christian and Egelman, Carolyn and Jaspan, Ciera and Murphy-Hill, Emerson},
  booktitle={2022 IEEE/ACM 44th International Conference on Software Engineering: Software Engineering in Society (ICSE-SEIS)},
  pages={41--55},
  year={2022},
  organization={IEEE},
  doi = {10.1109/ICSE-SEIS55304.2022.9793879},
  publisher = {IEEE},
  address = {USA}
}

@inproceedings{S14_novielli2022sensor,
  title={Sensor-based emotion recognition in software development: facial expressions as gold standard},
  author={Novielli, Nicole and Grassi, Daniela and Lanubile, Filippo and Serebrenik, Alexander},
  booktitle={2022 10th International Conference on Affective Computing and Intelligent Interaction (ACII)},
  pages={1--8},
  year={2022},
  organization={IEEE},
  doi = {10.1109/ACII55700.2022.9953808},
  publisher = {IEEE},
  address = {USA}
}

@article{S15_Sarker2023automated,
author = {Sarker, Jaydeb and Turzo, Asif Kamal and Dong, Ming and Bosu, Amiangshu},
title = {Automated Identification of Toxic Code Reviews Using ToxiCR},
year = {2023a},
issue_date = {September 2023},
publisher = {Association for Computing Machinery},
address = {New York, NY, USA},
volume = {32},
number = {5},
issn = {1049-331X},
url = {https://doi-org.proxy-ub.rug.nl/10.1145/3583562},
doi = {10.1145/3583562},
journal = {ACM Trans. Softw. Eng. Methodol.},
month = {jul},
articleno = {118},
numpages = {32}
}

@inproceedings{S16_Sarker2023,
  title={ToxiSpanSE: An Explainable Toxicity Detection in Code Review Comments},
  author={Sarker, Jaydeb and Sultana, Sayma and Wilson, Steven R and Bosu, Amiangshu},
  booktitle={2023 ACM/IEEE International Symposium on Empirical Software Engineering and Measurement (ESEM)},
  pages={1--12},
  year={2023b},
  organization={IEEE Computer Society},
  publisher = {IEEE},
  address = {USA}
}

@article{S17_trinkenreich2024predicting,
  title={Predicting Attrition among Software Professionals: Antecedents and Consequences of Burnout and Engagement},
  author={Trinkenreich, Bianca and Santos, Fabio and Stol, Klaas-Jan},
  journal={ACM Transactions on Software Engineering and Methodology},
  volume={33},
  number={8},
  pages={1--45},
  year={2024},
  publisher={ACM New York, NY},
  doi = {https://doi.org/10.1145/369162}
}

@article{S18_ferreira2024incivility,
  title={Incivility detection in open source code review and issue discussions},
  author={Ferreira, Isabella and Rafiq, Ahlaam and Cheng, Jinghui},
  journal={Journal of Systems and Software},
  volume={209},
  pages={111935},
  year={2024},
  publisher={Elsevier},
  doi = {https://doi.org/10.1016/j.jss.2023.111935}
}

@inproceedings{S19_imran2024emotion,
  title={Emotion Classification In Software Engineering Texts: A Comparative Analysis of Pre-trained Transformers Language Models},
  author={Imran, Mia Mohammad},
  booktitle={Proceedings of the Third ACM/IEEE International Workshop on NL-based Software Engineering},
  pages={73--80},
  year={2024},
  doi = {https://doi.org/10.1145/3643787.3648034},
  address = {USA},
  publisher = {ACM/IEEE}
}

@inproceedings{S20_singh2024softment,
  title={SOFTMENT: Detecting Mental Health and Wellbeing of Women in the Software Sector},
  author={Singh, Amrita and Anish, Preethu Rose and Ghaisas, Smita},
  booktitle={Companion of the 2024 on ACM International Joint Conference on Pervasive and Ubiquitous Computing},
  pages={405--411},
  year={2024},
  publisher = {Association for Computing Machinery},
  doi = {https://doi.org/10.1145/3675094.3678493},
  address = {ACM}
}

@inproceedings{S21_srikanteswara2024machine,
  title={Machine Learning-Based Stress Detection in IT Employees: A Data-Driven Approach for Workplace Well-being},
  author={Srikanteswara, Ramya and Rithicka, P and Kala, YV Shilpha and Rangaraj, Shreya and Devaiah, Varsha},
  booktitle={2024 Second International Conference on Data Science and Information System (ICDSIS)},
  pages={1--7},
  year={2024},
  organization={IEEE},
  doi = {10.1109/ICDSIS61070.2024.10594118},
  publisher = {IEEE},
  address = {USA}
}

@inproceedings{S22_ozakca2024artificial,
  title={Artificial Intelligence Based Employee Attrition Analysis and Prediction},
  author={Ozakca, Nazli Sila and Bulus, Ali and Cetin, Aydin},
  booktitle={2024 6th International Conference on Computing and Informatics (ICCI)},
  pages={512--517},
  year={2024},
  organization={IEEE},
  doi = {10.1109/ICCI61671.2024.10485157},
  address = {USA},
  publisher = {IEEE}
}

@inproceedings{S23_manikandan2024stress,
  title={Stress Monitoring with Computer Vision and Machine Learning for Software Employees},
  author={Manikandan, NK and Manivannan, D and Kavitha, M},
  booktitle={2024 2nd International Conference on Sustainable Computing and Smart Systems (ICSCSS)},
  pages={1016--1021},
  year={2024},
  organization={IEEE},
  doi = {10.1109/ICSCSS60660.2024.10625637},
  publisher = {IEEE},
  address = {USA}
}

@inproceedings{S24_Dovleac2021,
author="Dovleac, Raluca
and Risteiu, Marius
and Ionica, Andreea Cristina
and Leba, Monica",
editor="Rocha, {\'A}lvaro
and Adeli, Hojjat
and Dzemyda, Gintautas
and Moreira, Fernando
and Ramalho Correia, Ana Maria",
title="Mobile Burnout Estimation Device - An Agile Driven Pathway",
booktitle="World Conference on Information Systems and Technologies",
year="2021",
publisher="Springer International Publishing",
pages="522--531",
doi ="10.1007/978-3-030-72657-7_50",
address = "Germany"
}

@inproceedings{S26_geeth2024identification,
  title={Identification Emoticon Meaning using Facial Expression via Inception V3 as Well as KNN Methods},
  author={Geeth, MCS and Varma, M Krishna Satya and Sri, G Bhagya and Mythili, R and Naser, Sarmad Jaafar and Hassan, Enas},
  booktitle={2024 4th International Conference on Advance Computing and Innovative Technologies in Engineering (ICACITE)},
  pages={418--421},
  year={2024},
  organization={IEEE},
  doi = {10.1109/ICACITE60783.2024.10616678},
  publisher = {IEEE},
  address = {USA}
}

@article{S27_ballesteros2024facial,
  title={Facial emotion recognition through artificial intelligence},
  author={Ballesteros, Jes{\'u}s A and Ram{\'\i}rez V, Gabriel M and Moreira, Fernando and Solano, Andr{\'e}s and Pelaez, Carlos A},
  journal={Frontiers in Computer Science},
  volume={6},
  pages={1359471},
  year={2024},
  publisher={Frontiers Media SA},
  doi = {https://doi.org/10.3389/fcomp.2024.1359471}
}

@inproceedings{S28_awan2023creating,
  title={Creating Happier and More Productive Software Engineering Teams through AI and Machine Learning.},
  author={Awan, Wardah Naeem and Paasivaara, Maria and Gloor, Peter A and Salman, Iflaah},
  booktitle={ICSOB Companion},
  year={2023},
  publisher = {na},
  address = {na},
  pages = {na},
  numpages = {na}
}

@inproceedings{S29_jayathilake2023accurate,
  title={Accurate Stress Detection for Developers: Leveraging Low-Cost IoT Devices (ESP32 and MAX30102) to Analyze Heart Rate Variability via an External Mouse},
  author={Jayathilake, Amantha and Ranasinghe, Janudi and Perera, Dulshan and Bartholomeusz, Shehan and Rathnayake, Samadhi and Ganegoda, Devanshi},
  booktitle={2023 5th International Conference on Advancements in Computing (ICAC)},
  pages={89--94},
  year={2023},
  organization={IEEE},
  doi = {10.1109/ICAC60630.2023.10417345},
  publisher = {IEEE},
  address = {USA}
}

@inproceedings{S30_gamage2022machine,
  title={Machine learning approach to predict mental distress of IT Workforce in remote working environments},
  author={Gamage, Sanduni Nilushika and Asanka, PPG Dinesh},
  booktitle={2022 International Research Conference on Smart Computing and Systems Engineering (SCSE)},
  volume={5},
  pages={211--216},
  year={2022},
  organization={IEEE},
  doi = {10.1109/SCSE56529.2022.9905229},
  publisher = {IEE},
  address = {USA}
}

@article{S31_silva2023using,
  title={Using social media and personality traits to assess software developers’ emotional polarity},
  author={Silva, Leo and de Castro, Mar{\'\i}lia Gurgel and Silva, Miriam Bernardino and Santos, Milena and Kulesza, Uir{\'a} and Lima, Margarida and Madeira, Henrique},
  journal={PeerJ Computer Science},
  volume={9},
  pages={e1498},
  year={2023},
  publisher={PeerJ Inc.},
  doi = {https://doi.org/10.7717/peerj-cs.1498}
}

@inproceedings{S33_bleyl2022emotion,
  title={Emotion recognition on stackoverflow posts using bert},
  author={Bleyl, Donald and Buxton, Elham Khorasani},
  booktitle={2022 IEEE International Conference on Big Data (Big Data)},
  pages={5881--5885},
  year={2022},
  organization={IEEE},
  doi = {10.1109/BigData55660.2022.10020161},
  publisher = {IEEE},
  address = {USA}
}

@inproceedings{S34_maheshwarkar2021analysis,
  title={Analysis of written interactions in open-source communities using RCNN},
  author={Maheshwarkar, Aashay and Kumar, Animesh and Gupta, Manas},
  booktitle={2021 3rd International conference on advances in computing, communication control and networking (ICAC3N)},
  pages={996--1001},
  year={2021},
  organization={IEEE},
  doi = {10.1109/ICAC3N53548.2021.9725610},
  address = {USA},
  publisher = {IEEE}
}

@conference{S37_Garcia2013,  author={Garcia, David and Zanetti, Marcelo Serrano and Schweitzer, Frank},  booktitle={2013 International Conference on Cloud and Green Computing},   title={The Role of Emotions in Contributors Activity: A Case Study on the GENTOO Community},   year={2013},  volume={},  number={},  pages={410-417},  doi={10.1109/CGC.2013.71},
publisher={IEEE}, address = {USA}}

@conference{S38_Muller2015,  author={Müller, Sebastian C. and Fritz, Thomas},  booktitle={2015 IEEE/ACM 37th IEEE Intl. Conf. on Soft. Eng.},   title={Stuck and Frustrated or in Flow and Happy: Sensing Developers' Emotions and Progress},   year={2015},  volume={1},  number={},  pages={688-699},  doi={10.1109/ICSE.2015.334},
publisher = {IEEE/ACM}, address = {USA}}

@inproceedings{S39_Pletea2014,
author = {Pletea, Daniel and Vasilescu, Bogdan and Serebrenik, Alexander},
title = {Security and Emotion: Sentiment Analysis of Security Discussions on GitHub},
year = {2014},
isbn = {9781450328630},
publisher = {Association for Computing Machinery},
address = {New York, NY, USA},
doi = {10.1145/2597073.2597117},
booktitle = {Proceedings of the 11th Working Conference on Mining Software Repositories},
pages = {348–351},
numpages = {4},
location = {Hyderabad, India},
series = {MSR 2014}
}

@conference{S40_ortu2016arsonists,
  title={Arsonists or firefighters? Affectiveness in agile software development},
  author={Ortu, Marco and Destefanis, Giuseppe and Counsell, Steve and Swift, Stephen and Tonelli, Roberto and Marchesi, Michele},
  booktitle={International Conference on Agile Software Development},
  pages={144--155},
  year={2016},
  organization={Springer International Publishing},
  doi = {10.1007/978-3-319-33515-5 12},
  address = {Germany},
  publisher = {Springer}
}

@inproceedings{S41_Sinha2016,
author = {Sinha, Vinayak and Lazar, Alina and Sharif, Bonita},
title = {Analyzing Developer Sentiment in Commit Logs},
year = {2016},
isbn = {9781450341868},
publisher = {Association for Computing Machinery},
address = {New York, NY, USA},
doi = {10.1145/2901739.2903501},
booktitle = {Proceedings of the 13th International Conference on Mining Software Repositories},
pages = {520–523},
numpages = {4},
location = {Austin, Texas},
series = {MSR '16}
}

@conference{S42_Nogueira2013,  author={Nogueira, Pedro A. and Rodrigues, Rui and Oliveira, Eugénio and Nacke, Lennart E.},  booktitle={2013 IEEE/WIC/ACM Intl Joint Conferences on Web Intelligence (WI) and Intelligent Agent Technologies (IAT)},   title={A Hybrid Approach at Emotional State Detection: Merging Theoretical Models of Emotion with Data-Driven Statistical Classifiers},   year={2013},  volume={2},  number={},  pages={253-260},  doi={10.1109/WI-IAT.2013.117}, 
    address = {New York, USA},
    publisher = {IEEE}}

@inproceedings{S43_radevski2015real,
  title={Real-time monitoring of neural state in assessing and improving software developers' productivity},
  author={Radevski, Stevche and Hata, Hideaki and Matsumoto, Kenichi},
  booktitle={2015 IEEE/ACM 8th International Workshop on Cooperative and Human Aspects of Software Engineering},
  pages={93--96},
  year={2015},
  organization={IEEE},
  doi = {10.1109/CHASE.2015.28},
  address = {USA},
  publisher = {IEEE}
}

@article{S44_androutsou2023automated,
  title={Automated multimodal stress detection in computer office workspace},
  author={Androutsou, Thelma and Angelopoulos, Spyridon and Hristoforou, Evangelos and Matsopoulos, George K and Koutsouris, Dimitrios D},
  journal={Electronics},
  volume={12},
  number={11},
  pages={2528},
  year={2023},
  publisher={MDPI},
  doi = { https://doi.org/10.3390/electronics12112528}
}

@ARTICLE{S46_GirardiLanubile2021,  author={Girardi, Daniela and Lanubile, Filippo and Novielli, Nicole and Serebrenik, Alexander},  journal={IEEE Transactions on Software Engineering},   title={Emotions and Perceived Productivity of Software Developers at the Workplace},   year={2021},  volume={48},  number={9},  pages={1-1},  doi={10.1109/TSE.2021.3087906}}

@inproceedings{S47_sarker2022identification,
  title={Identification and mitigation of toxic communications among open source software developers},
  author={Sarker, Jaydeb},
  booktitle={Proceedings of the 37th IEEE/ACM International Conference on Automated Software Engineering},
  pages={1--5},
  year={2022},
  doi = {https://doi.org/10.1145/3551349.355957},
  address = {USA},
  publisher = {IEEE/ACM}
}

@inproceedings{S48_islam2019marvalous,
  title={MarValous: Machine learning based detection of emotions in the valence-arousal space in software engineering text},
  author={Islam, Md Rakibul and Ahmmed, Md Kauser and Zibran, Minhaz F},
  booktitle={Proceedings of the 34th ACM/SIGAPP Symposium on Applied Computing},
  pages={1786--1793},
  year={2019},
  doi = {10.1145/3297280.3297455},
  publisher = {ACM},
  address = {USA}
}

@conference{S51_tourani2014monitoring,  author={Tourani, Parastou and Jiang, Yujuan and Adams, Bram},  booktitle={2014 International on Computer Science and Software Engineering (CASCON)},   title={Monitoring sentiment in open source mailing lists: exploratory study on the apache ecosystem.},   year={2014},  volume={14},  number={},  pages={34-44},
publisher = {ACM}, address = {USA}}

@inproceedings{S52_reddy2018machine,
  title={Machine learning techniques for stress prediction in working employees},
  author={Reddy, U Srinivasulu and Thota, Aditya Vivek and Dharun, A},
  booktitle={2018 IEEE International Conference on Computational Intelligence and Computing Research (ICCIC)},
  pages={1--4},
  year={2018},
  organization={IEEE},
  doi = {10.1109/ICCIC.2018.8782395},
  publisher = {IEEE},
  address = {USA}
}

@inproceedings{S55_kolakowska2013emotion,
  title={Emotion recognition and its application in software engineering},
  author={Ko{\l}akowska, Agata and Landowska, Agnieszka and Szwoch, Mariusz and Szwoch, Wioleta and Wr{\'o}bel, Micha{\l} R},
  booktitle={2013 6th International Conference on Human System Interactions (HSI)},
  pages={532--539},
  year={2013},
  organization={IEEE},
  doi = {10.1109/HSI.2013.6577877},
  publisher = {IEEE},
  address = {USA}
}

@inproceedings{S60_mishra2024exploring,
  title={Exploring chatgpt for toxicity detection in github},
  author={Mishra, Shyamal and Chatterjee, Preetha},
  booktitle={Proceedings of the 2024 ACM/IEEE 44th International Conference on Software Engineering: New Ideas and Emerging Results},
  pages={6--10},
  year={2024},
  doi = {https://doi.org/10.1145/3639476.3639777},
  publisher = {ACM/IEEE},
  address = {USA}
}

@article{S63_cabrera2020classifying,
  title={Classifying emotions in Stack Overflow and JIRA using a multi-label approach},
  author={Cabrera-Diego, Luis Adri{\'a}n and Bessis, Nik and Korkontzelos, Ioannis},
  journal={Knowledge-Based Systems},
  volume={195},
  pages={105633},
  year={2020},
  publisher={Elsevier},
  doi = {https://doi.org/10.1016/j.knosys.2020.105633}
}

@article{S68_wagan2025multilabeled,
  title={Multilabeled Emotions Classification in Software Engineering Text Using Convolutional Neural Networks and Word Embeddings},
  author={Wagan, Atif Ali and Li, Shuaiyong},
  journal={Journal of Software: Evolution and Process},
  volume={37},
  number={3},
  pages={e70010},
  year={2025},
  publisher={Wiley Online Library},
  doi = { https://doi.org/10.1002/smr.70010}
}

@conference{S70_Sarker2020,  author={Sarker, Jaydeb and Turzo, Asif Kamal and Bosu, Amiangshu},  booktitle={2020 27th Asia-Pacific Software Engineering Conference (APSEC)},   title={A Benchmark Study of the Contemporary Toxicity Detectors on Software Engineering Interactions},   year={2020},  volume={},  number={},  pages={218-227},  doi={10.1109/APSEC51365.2020.00030}, address = {USA}, publisher = {ACM}}

@inproceedings{S71_bhat2021say,
  title={Say ‘YES’to positivity: Detecting toxic language in workplace communications},
  author={Bhat, Meghana Moorthy and Hosseini, Saghar and Hassan, Ahmed and Bennett, Paul and Li, Weisheng},
  booktitle={Findings of the Association for Computational Linguistics: EMNLP 2021},
  pages={2017--2029},
  year={2021},
  doi = {10.18653/v1/2021.findings-emnlp.173},
  address = {na},
  publisher = {ACL Anthology}
}

@article{S72_yang2021behavioral,
  title={Behavioral and physiological signals-based deep multimodal approach for mobile emotion recognition},
  author={Yang, Kangning and Wang, Chaofan and Gu, Yue and Sarsenbayeva, Zhanna and Tag, Benjamin and Dingler, Tilman and Wadley, Greg and Goncalves, Jorge},
  journal={IEEE Transactions on Affective Computing},
  volume={14},
  number={2},
  pages={1082--1097},
  year={2021},
  publisher={IEEE},
  doi = {10.1109/TAFFC.2021.3100868}
}

@article{S73_rahman2024words,
  title={Do words have power? understanding and fostering civility in code review discussion},
  author={Rahman, Md Shamimur and Codabux, Zadia and Roy, Chanchal K},
  journal={Proceedings of the ACM on Software Engineering},
  volume={1},
  number={FSE},
  pages={1632--1655},
  year={2024},
  publisher={ACM New York, NY, USA},
  doi = {https://doi.org/10.1145/3660780}
}

@article{S74_booth2022toward,
  title={Toward robust stress prediction in the age of wearables: Modeling perceived stress in a longitudinal study with information workers},
  author={Booth, Brandon M and Vrzakova, Hana and Mattingly, Stephen M and Martinez, Gonzalo J and Faust, Louis and D’Mello, Sidney K},
  journal={IEEE Transactions on Affective Computing},
  volume={13},
  number={4},
  pages={2201--2217},
  year={2022},
  publisher={IEEE},
  doi = {10.1109/TAFFC.2022.3188006}
}

@inproceedings{S75_padha2022quantum,
  title={Quantum enhanced machine learning for unobtrusive stress monitoring},
  author={Padha, Anupama and Sahoo, Anita},
  booktitle={Proceedings of the 2022 Fourteenth International Conference on Contemporary Computing},
  pages={476--483},
  year={2022},
  doi = {https://doi.org/10.1145/3549206.3549288},
  publisher = {ACM},
  address = {New York, USA}
}

@article{S77_vrzakova2020affect,
  title={Affect recognition in code review: An in-situ biometric study of reviewer’s affect},
  author={Vrzakova, Hana and Begel, Andrew and Meht{\"a}talo, Lauri and Bednarik, Roman},
  journal={Journal of Systems and Software},
  volume={159},
  pages={110434},
  year={2020},
  publisher={Elsevier},
  doi = {https://doi.org/10.1016/j.jss.2019.110434}
}

@article{S78_rissler2020or,
  title={To be or not to be in flow at work: physiological classification of flow using machine learning},
  author={Rissler, Raphael and Nadj, Mario and Li, Maximilian X and Loewe, Nico and Knierim, Michael T and Maedche, Alexander},
  journal={IEEE transactions on affective computing},
  volume={14},
  number={1},
  pages={463--474},
  year={2020},
  publisher={IEEE},
  doi = {10.1109/TAFFC.2020.3045269}
}

@inproceedings{S79_fritz2016leveraging,
  title={Leveraging biometric data to boost software developer productivity},
  author={Fritz, Thomas and M{\"u}ller, Sebastian C},
  booktitle={2016 IEEE 23rd international conference on software analysis, evolution, and reengineering (SANER)},
  volume={5},
  pages={66--77},
  year={2016},
  organization={IEEE},
  doi = {10.1109/SANER.2016.107},
  publisher = {IEEE},
  address = {USA}
}

@inproceedings{S81_anany2019influence,
  title={Influence of Emotions on Software Developer Productivity.},
  author={Anany, Mohammed R and Hussien, Heba and Aly, Sherif G and Sakr, Nourhan},
  booktitle={PECCS},
  pages={75--82},
  year={2019},
  publisher = {scitepress},
  address = {Portugal},
  doi = {10.5220/0008068800750082}
}

@inproceedings{S82_nogueira2015modelling,
  title={Modelling human emotion in interactive environments: Physiological ensemble and grounded approaches for synthetic agents},
  author={Nogueira, Pedro A and Rodrigues, Rui and Oliveira, Eug{\'e}nio and Nacke, Lennart E},
  booktitle={Web Intelligence},
  volume={13},
  pages={195--214},
  year={2015},
  organization={SAGE Publications Sage UK: London, England},
  doi = {https://doi.org/10.3233/WEB-150321},
  publisher = {Sage},
  address = {London, England}
}

@article{S83_naegelin2023interpretable,
  title={An interpretable machine learning approach to multimodal stress detection in a simulated office environment},
  author={Naegelin, Mara and Weibel, Raphael P and Kerr, Jasmine I and Schinazi, Victor R and La Marca, Roberto and von Wangenheim, Florian and Hoelscher, Christoph and Ferrario, Andrea},
  journal={Journal of biomedical informatics},
  volume={139},
  pages={104299},
  year={2023},
  publisher={Elsevier},
  doi = {https://doi.org/10.1016/j.jbi.2023.104299}
}

@inproceedings{S84_epp2011identifying,
  title={Identifying emotional states using keystroke dynamics},
  author={Epp, Clayton and Lippold, Michael and Mandryk, Regan L},
  booktitle={Proceedings of the sigchi conference on human factors in computing systems},
  pages={715--724},
  year={2011},
  publisher = {Association Computer Machinery},
  doi = {https://doi.org/10.1145/1978942.1979046},
  address = {New York, USA}
}

@article{S85_munoz2022text,
  title={A text classification approach to detect psychological stress combining a lexicon-based feature framework with distributional representations},
  author={Mu{\~n}oz, Sergio and Iglesias, Carlos A},
  journal={Information Processing \& Management},
  volume={59},
  number={5},
  pages={103011},
  year={2022},
  publisher={Elsevier},
  doi = {https://doi.org/10.1016/j.ipm.2022.103011}
}

@article{S86_koldijk2016detecting,
  title={Detecting work stress in offices by combining unobtrusive sensors},
  author={Koldijk, Saskia and Neerincx, Mark A and Kraaij, Wessel},
  journal={IEEE Transactions on affective computing},
  volume={9},
  number={2},
  pages={227--239},
  year={2016},
  publisher={IEEE},
  doi = { 10.1109/TAFFC.2016.2610975}
}

@article{S87_carneiro2012multimodal,
  title={Multimodal behavioral analysis for non-invasive stress detection},
  author={Carneiro, Davide and Castillo, Jos{\'e} Carlos and Novais, Paulo and Fern{\'a}ndez-Caballero, Antonio and Neves, Jos{\'e}},
  journal={Expert Systems with Applications},
  volume={39},
  number={18},
  pages={13376--13389},
  year={2012},
  publisher={Elsevier},
  doi = {https://doi.org/10.1016/j.eswa.2012.05.065}
}

@article{S88_pepa2020stress,
  title={Stress detection in computer users from keyboard and mouse dynamics},
  author={Pepa, Lucia and Sabatelli, Antonio and Ciabattoni, Lucio and Monteri{\`u}, Andrea and Lamberti, Fabrizio and Morra, Lia},
  journal={IEEE Transactions on Consumer Electronics},
  volume={67},
  number={1},
  pages={12--19},
  year={2020},
  publisher={IEEE},
  doi = {10.1109/TCE.2020.3045228}
}

@article{S89_alberdi2018using,
  title={Using smart offices to predict occupational stress},
  author={Alberdi, Ane and Aztiria, Asier and Basarab, Adrian and Cook, Diane J},
  journal={International Journal of Industrial Ergonomics},
  volume={67},
  pages={13--26},
  year={2018},
  publisher={Elsevier},
  doi = {https://doi.org/10.1016/j.ergon.2018.04.005}
}

@conference{S91_klunder2020identifying,
  title={Identifying the mood of a software development team by analyzing text-based communication in chats with machine learning},
  author={Kl{\"u}nder, Jil and Horstmann, Julian and Karras, Oliver},
  booktitle={International Conference on Human-Centred Software Engineering},
  pages={133--151},
  year={2020},
  organization={Springer},
  doi = {10.1007/978-3-030-64266-2_8},
  publisher = {Springer},
  address = {The Netherlands}
}

@article{S92_nath2021burnoutwords,
  title={Burnoutwords-detecting burnout for a clinical setting},
  author={Nath, Sukanya and Kurpicz-Briki, Mascha and Kurpicz-Briki, SN and Nath, SM and Kurpicz-Briki, M},
  journal={Comput Sci Inf Technol},
  volume={11},
  pages={177},
  year={2021},
  doi = {10.5121/csit.2021.111815}
}

@article{S93_vizer2009automated,
  title={Automated stress detection using keystroke and linguistic features: An exploratory study},
  author={Vizer, Lisa M and Zhou, Lina and Sears, Andrew},
  journal={International Journal of Human-Computer Studies},
  volume={67},
  number={10},
  pages={870--886},
  year={2009},
  publisher={Elsevier},
  doi = {10.1016/j.ijhcs.2009.07.005}
}

@article{kitchenham2004procedures,
  title={Procedures for performing systematic reviews},
  author={Kitchenham, Barbara},
  journal={Keele, UK, Keele University},
  volume={33},
  number={2004},
  pages={1--26},
  year={2004}
}

@article{upchurch2001using,
  title={Using card sorts to elicit web page quality attributes},
  author={Upchurch, Linda and Rugg, Gordon and Kitchenham, Barbara},
  journal={Ieee software},
  volume={18},
  number={4},
  pages={84},
  year={2001},
  publisher={IEEE Computer Society},
  doi = {10.1109/MS.2001.936222}
}

@article{maslach2001job,
  title={Job burnout},
  author={Maslach, Christina and Schaufeli, Wilmar B and Leiter, Michael P},
  journal={Annual review of psychology},
  volume={52},
  number={1},
  pages={397--422},
  year={2001},
  publisher={Annual Reviews 4139 El Camino Way, PO Box 10139, Palo Alto, CA 94303-0139, USA}
}

@book{maslach1997maslach,
  title={Maslach burnout inventory},
  author={Maslach, Christina and Jackson, Susan E and Leiter, Michael P},
  year={1997},
  publisher={Scarecrow Education},
  address = {USA}
}

@article{P36sonnentag1994stressor,
  title={Stressor-burnout relationship in software development teams},
  author={Sonnentag, Sabine and Brodbeck, Felix C and Heinbokel, Torsten and Stolte, Wolfgang},
  journal={Journal of occupational and organizational psychology},
  volume={67},
  number={4},
  pages={327--341},
  year={1994},
  publisher={Wiley Online Library},
  doi = {10.1111/j.2044-8325.1994.tb00571.x}
}

@article{martin2018google,
  title={Google Scholar, Web of Science, and Scopus: A systematic comparison of citations in 252 subject categories},
  author={Mart{\'\i}n-Mart{\'\i}n, Alberto and Orduna-Malea, Enrique and Thelwall, Mike and L{\'o}pez-C{\'o}zar, Emilio Delgado},
  journal={Journal of informetrics},
  volume={12},
  number={4},
  pages={1160--1177},
  year={2018},
  publisher={Elsevier},
  doi = {10.1016/j.joi.2018.09.002}
}

@conference{P72kaur2022didn,
  title={“I Didn’t Know I Looked Angry”: Characterizing Observed Emotion and Reported Affect at Work},
  author={Kaur, Harmanpreet and McDuff, Daniel and Williams, Alex C and Teevan, Jaime and Iqbal, Shamsi T},
  booktitle={CHI Conference on Human Factors in Computing Systems},
  pages={1--18},
  year={2022},
  doi = {10.1145/3491102.3517453},
  publisher = {ACM},
  address = {New York, USA}
}

@misc{Kitchenham07guidelinesfor,
    author = {B. Kitchenham and S Charters},
    title = {Guidelines for performing Systematic Literature Reviews in Software Engineering},
    year = {2007},
    publisher={Technical report, ver. 2.3 ebse technical report. ebse}
}

@article{brereton2007lessons,
  title={Lessons from applying the systematic literature review process within the software engineering domain},
  author={Brereton, Pearl and Kitchenham, Barbara A and Budgen, David and Turner, Mark and Khalil, Mohamed},
  journal={Journal of systems and software},
  volume={80},
  number={4},
  pages={571--583},
  year={2007},
  publisher={Elsevier},
  doi = {10.1016/j.jss.2006.07.009}
}

@conference{yordanova2021agile,
  title={Agile Application for Innovation Projects in Science Organizations-Knowledge Gap and State of Art},
  author={Yordanova, Zornitsa},
  booktitle={International Conference on Information Technology \& Systems},
  pages={108--117},
  year={2021},
  organization={Springer},
  address = {Germany},
  publisher = {Springer}
}

@article{lin2022opinion,
  title={Opinion mining for software development: a systematic literature review},
  author={Lin, Bin and Cassee, Nathan and Serebrenik, Alexander and Bavota, Gabriele and Novielli, Nicole and Lanza, Michele},
  journal={ACM Transactions on Software Engineering and Methodology (TOSEM)},
  volume={31},
  number={3},
  pages={1--41},
  year={2022},
  publisher={ACM New York, NY},
  doi = {10.1145/3490388}
}

@article{obaidi2021development,
  title={Development and application of sentiment analysis tools in software engineering: A systematic literature review},
  author={Obaidi, Martin and Kl{\"u}nder, Jil},
  journal={Evaluation and Assessment in Software Engineering},
  pages={80--89},
  year={2021},
  doi = {10.1145/3463274.3463328},
  number = {na},
  volume = {na}
}

@article{sanchez2019taking,
  title={Taking the emotional pulse of software engineering—A systematic literature review of empirical studies},
  author={S{\'a}nchez-Gord{\'o}n, Mary and Colomo-Palacios, Ricardo},
  journal={Information and Software Technology},
  volume={115},
  pages={23--43},
  year={2019},
  publisher={Elsevier},
  doi = {10.1016/j.infsof.2019.08.002}
}

@article{tawsif2022systematic,
  title={A Systematic Review on Emotion Recognition System Using Physiological Signals: Data Acquisition and Methodology},
  author={Tawsif, K and Aziz, Nor Azlina Ab and Raja, J Emerson and Hossen, J and Jesmeen, MZH},
  journal={Emerging Science Journal},
  volume={6},
  number={5},
  pages={1167--1198},
  year={2022},
  doi = {10.28991/ESJ-2022-06-05-017}
}

@article{zucco2020sentiment,
  title={Sentiment analysis for mining texts and social networks data: Methods and tools},
  author={Zucco, Chiara and Calabrese, Barbara and Agapito, Giuseppe and Guzzi, Pietro H and Cannataro, Mario},
  journal={Wiley Interdisciplinary Reviews: Data Mining and Knowledge Discovery},
  volume={10},
  number={1},
  pages={e1333},
  year={2020},
  publisher={Wiley Online Library},
  doi = {10.1002/widm.1333}
}

@article{singh2022cognitive,
  title={Cognitive Computing in Mental Healthcare: a Review of Methods and Technologies for Detection of Mental Disorders},
  author={Singh, Jaiteg and Hamid, Mir Aamir},
  journal={Cognitive Computation},
  volume={14},
  number={6},
  pages={2169--2186},
  year={2022},
  publisher={Springer},
  doi = {10.1007/s12559-022-10042-2}
}

@article{sardar2022systematic,
  title={A Systematic Literature Review on Machine Learning Algorithms for Human Status Detection},
  author={Sardar, Suman Kalyan and Kumar, Naveen and Lee, Seul Chan},
  journal={IEEE Access},
  volume={10},
  pages={74366--74382},
  year={2022},
  publisher={IEEE},
  doi = {10.1109/ACCESS.2022.3190967}
}

@article{thelwall2010sentiment,
  title={Sentiment strength detection in short informal text},
  author={Thelwall, Mike and Buckley, Kevan and Paltoglou, Georgios and Cai, Di and Kappas, Arvid},
  journal={Journal of the American society for information science and technology},
  volume={61},
  number={12},
  pages={2544--2558},
  year={2010},
  publisher={Wiley Online Library},
  doi = {10.1002/asi.21416}
}

@article{paullada2021data,
  title={Data and its (dis) contents: A survey of dataset development and use in machine learning research},
  author={Paullada, Amandalynne and Raji, Inioluwa Deborah and Bender, Emily M and Denton, Emily and Hanna, Alex},
  journal={Patterns},
  volume={2},
  number={11},
  pages={100336},
  year={2021},
  publisher={Elsevier},
  doi = {10.1016/j.patter.2021.100336}
}

@inproceedings{liebchen2008data,
  title={Data sets and data quality in software engineering},
  author={Liebchen, Gernot A and Shepperd, Martin},
  booktitle={Proceedings of the 4th international workshop on Predictor models in software engineering},
  pages={39--44},
  volume={},
  number={},
  year={2008},
  doi = {10.1145/1370788.1370799},
  publisher = {ACM},
  address = {USA}
}

@misc{nitesh2019data,
  title={Data Quality Model for Machine learning},
  author={{Nitesh Varma Rudraraju, Nitesh and Varun Boyanapally, Varun}},
  year={2019},
  howpublished={https://urn.kb.se/resolve?urn=urn:nbn:se:bth-18498},
  publisher="n/a",
  note = {Online; February 2023}
}

@article{schoeps2019effects,
  title={Effects of emotional skills training to prevent burnout syndrome in schoolteachers},
  author={Schoeps, Konstanze and Tamarit, Alicia and de la Barrera, Usue and Barr{\'o}n, Remedios Gonz{\'a}lez},
  journal={Ansiedad y Estr{\'e}s},
  volume={25},
  number={1},
  pages={7--13},
  year={2019},
  publisher={Elsevier},
  doi = {10.1016/j.anyes.2019.01.002}
}

@article{alessandri2018job,
  title={Job burnout: The contribution of emotional stability and emotional self-efficacy beliefs},
  author={Alessandri, Guido and Perinelli, Enrico and De Longis, Evelina and Schaufeli, Wilmar B and Theodorou, Annalisa and Borgogni, Laura and Caprara, Gian Vittorio and Cinque, Luigi},
  journal={Journal of occupational and organizational psychology},
  volume={91},
  number={4},
  pages={823--851},
  year={2018},
  publisher={Wiley Online Library},
  doi = {10.1111/joop.12225}
}

@article{holmqvist2006burnout,
  title={Burnout and psychiatric staff's feelings towards patients},
  author={Holmqvist, Rolf and Jeanneau, Madeleine},
  journal={Psychiatry research},
  volume={145},
  number={2-3},
  pages={207--213},
  year={2006},
  publisher={Elsevier},
  doi = {10.1016/j.psychres.2004.08.012}
}

@article{liu2021negative,
  title={Negative emotions and job burnout in news media workers: a moderated mediation model of rumination and empathy},
  author={Liu, Mingxiao and Wang, Ning and Wang, Pengcheng and Wu, Haomeng and Ding, Xianger and Zhao, Fengqing},
  journal={Journal of Affective Disorders},
  volume={279},
  pages={75--82},
  year={2021},
  publisher={Elsevier},
  doi = {10.1016/j.jad.2020.09.123}
}

@article{abellanoza2018burnout,
  title={Burnout in ER nurses: Review of the literature and interview themes},
  author={Abellanoza, Adrian and Provenzano-Hass, Nicolette and Gatchel, Robert J},
  journal={Journal of Applied Biobehavioral Research},
  volume={23},
  number={1},
  pages={e12117},
  year={2018},
  publisher={Wiley Online Library},
  doi = {10.1111/jabr.12117}
}

@article{garcia2023feature,
  title={Feature engineering of EEG applied to mental disorders: a systematic mapping study},
  author={Garc{\'\i}a-Ponsoda, Sandra and Garc{\'\i}a-Carrasco, Jorge and Teruel, Miguel A and Mat{\'e}, Alejandro and Trujillo, Juan},
  journal={Applied Intelligence},
  volume={53},
  number={na},  
  pages={1--41},
  year={2023},
  publisher={Springer},
  doi = {10.1007/s10489-023-04702-5}
}

@online{SurveyHaystack,
  author = {Junaedi Ali},
  title = {83\% of Developers Suffer From Burnout, Haystack Analytics Study Finds},
  year = 2021,
  url = {https://www.usehaystack.io/blog/83-of-developers-suffer-from-burnout-haystack-analytics-study-finds},
  urldate = {2021-07-09},
  note = {last accessed, 19th of November 2023},
  organization = {Haystack}
}

@online{SurveyCodeAhoy,
  author = {Umer Mansoor},
  title = {Burnout in Software Development - Survey Results 2021},
  year = 2021,
  url = {https://codeahoy.com/2021/10/01/software-developer-burn-out-survey/},
  urldate = {2021-10-01},
  note = {last accessed, 19th of November 2023},
  organization  ={Codeahoy}
}

@article{constantino2023perceptions,
  title={Perceptions of open-source software developers on collaborations: An interview and survey study},
  author={Constantino, Kattiana and Souza, Mauricio and Zhou, Shurui and Figueiredo, Eduardo and K{\"a}stner, Christian},
  journal={Journal of Software: Evolution and Process},
  volume={35},
  number={5},
  pages={e2393},
  year={2023},
  publisher={Wiley Online Library}
}

@article{ratanawongsa2008physician,
  title={Physician burnout and patient-physician communication during primary care encounters},
  author={Ratanawongsa, Neda and Roter, Debra and Beach, Mary Catherine and Laird, Shivonne L and Larson, Susan M and Carson, Kathryn A and Cooper, Lisa A},
  journal={Journal of general internal medicine},
  volume={23},
  pages={1581--1588},
  year={2008},
  publisher={Springer}
}

@article{robbins2019provider,
  title={Provider burnout and patient-provider communication in the context of hypertension care},
  author={Robbins, Rebecca and Butler, Mark and Schoenthaler, Antoinette},
  journal={Patient Education and Counseling},
  volume={102},
  number={8},
  pages={1452--1459},
  year={2019},
  publisher={Elsevier}
}

@article{miller2007compassionate,
  title={Compassionate communication in the workplace: Exploring processes of noticing, connecting, and responding},
  author={Miller, Katherine I},
  journal={Journal of Applied Communication Research},
  volume={35},
  number={3},
  pages={223--245},
  year={2007},
  publisher={Taylor \& Francis}
}

@article{akhavan2022going,
  title={“Going through the motions”: A qualitative exploration of the impact of emergency medicine resident burnout on patient care},
  author={Akhavan, Arvin Radfar and Strout, Tania D and Germann, Carl A and Nelson, Sara W and Jauregui, Joshua and Lu, Dave W},
  journal={AEM education and training},
  volume={6},
  number={5},
  pages={e10809},
  year={2022},
  publisher={Wiley Online Library}
}

@article{li2017roles,
  title={The roles of the talent development environment on athlete burnout: a qualitative study.},
  author={Li, ChunXiao and Wang, CheeKeng and Pyun, DoYoung and others},
  journal={International Journal of Sport Psychology},
  volume={48},
  number={2},
  pages={143--164},
  year={2017},
  publisher={Edizioni Luigi Pozzi}
}

@article{hesham2023special,
  title={Special education teachers’ mental health after reopening schools during Covid-19},
  author={Hesham Abdou Ahmed, Gelan},
  journal={Plos one},
  volume={18},
  number={5},
  pages={e0284870},
  year={2023},
  publisher={Public Library of Science San Francisco, CA USA}
}

@article{salles2020assessment,
  title={Assessment of psychological capital at work by physiotherapists},
  author={Salles, Fagner Luiz Pacheco and d'Angelo, Marcia Juliana},
  journal={Physiotherapy Research International},
  volume={25},
  number={3},
  pages={e1828},
  year={2020},
  publisher={Wiley Online Library}
}

@article{Lee2016jpts2016286,
  title={Impact of work environment and work-related stress on turnover intention in          physical therapists},
  author={Byoung-kwon Lee and Dong-kwon Seo and Jang-Tae Lee and A-Ram Lee and Ha-Neul Jeon and Dong-Uk Han},
  journal={Journal of Physical Therapy Science},
  volume={28},
  number={8},
  pages={2358-2361},
  year={2016},
  doi={10.1589/jpts.28.2358}
}

@article{hall2011systematic,
  title={A systematic literature review on fault prediction performance in software engineering},
  author={Hall, Tracy and Beecham, Sarah and Bowes, David and Gray, David and Counsell, Steve},
  journal={IEEE Transactions on Software Engineering},
  volume={38},
  number={6},
  pages={1276--1304},
  year={2011},
  publisher={IEEE}
}

@article{ismail2023systematic,
  title={A systematic review of emotion recognition using cardio-based signals},
  author={Ismail, Sharifah Noor Masidayu Sayed and Aziz, Nor Azlina Ab and Ibrahim, Siti Zainab and Mohamad, Mohd Saberi},
  journal={ICT Express},
  year={2023},
  volume={10},
  number={1},
  publisher={Elsevier},
  pages={156--183},
  address = {The Netherlands}
}

@article{vaswani2017attention,
  title={Attention is all you need},
  author={Vaswani, Ashish and Shazeer, Noam and Parmar, Niki and Uszkoreit, Jakob and Jones, Llion and Gomez, Aidan N and Kaiser, {\L}ukasz and Polosukhin, Illia},
  journal={Advances in neural information processing systems},
  volume={30},
  pages = {na},
  numpages = {na},
  year={2017}
}

@inproceedings{kaur2020optimizing,
  title={Optimizing for happiness and productivity: Modeling opportune moments for transitions and breaks at work},
  author={Kaur, Harmanpreet and Williams, Alex C and McDuff, Daniel and Czerwinski, Mary and Teevan, Jaime and Iqbal, Shamsi T},
  booktitle={Proceedings of the 2020 CHI Conference on Human Factors in Computing Systems},
  pages={1--15},
  year={2020},
  publisher = {ACM},
  address = {New York, USA}
}

@article{ghosh2021depression,
  title={Depression intensity estimation via social media: a deep learning approach},
  author={Ghosh, Shreya and Anwar, Tarique},
  journal={IEEE Transactions on Computational Social Systems},
  volume={8},
  number={6},
  pages={1465--1474},
  year={2021},
  publisher={IEEE}
}

@article{ahmed2022machine,
  title={Machine learning models to detect anxiety and depression through social media: A scoping review},
  author={Ahmed, Arfan and Aziz, Sarah and Toro, Carla T and Alzubaidi, Mahmood and Irshaidat, Sara and Serhan, Hashem Abu and Abd-Alrazaq, Alaa A and Househ, Mowafa},
  journal={Computer Methods and Programs in Biomedicine Update},
  pages={100066},
  year={2022},
  number = {na},
  volume={2},
  publisher={Elsevier}
}

@misc{shenouda2023improving,
  title={Improving Bug Assignment and Developer Allocation in Soft-ware Engineering through Interpretable Machine Learning Models},
  author={Shenouda, Mina Samir and Sherief, Nada H and Abdelmoez, Walid M},
  year={2023},
  publisher={Preprints}
}

@article{samir2023improving,
  title={Improving Bug Assignment and Developer Allocation in Software Engineering through Interpretable Machine Learning Models},
  author={Samir, Mina and Sherief, Nada and Abdelmoez, Walid},
  journal={Computers},
  volume={12},
  number={7},
  pages={128},
  year={2023},
  publisher={MDPI}
}

@misc{McKinsey2023,
    author =  {Chui, Michael and Roberts, Roger and Rodchenko, Tanya and Singla, Alex and Sukharevsky, Alex and Lareina, Yee and Zurkiya, Delphine},
    title = {What every CEO should know about generative AI},
    year={2023},
    howpublished={https://www.mckinsey.com/capabilities/mckinsey-digital/our-insights/what-every-ceo-should-know-about-generative-ai},
    publisher="McKinsey Digital",
    note = {Online; February 2023}
}

@article{mendes2017privacy,
  title={Privacy-preserving data mining: methods, metrics, and applications},
  author={Mendes, Ricardo and Vilela, Jo{\~a}o P},
  journal={IEEE Access},
  volume={5},
  pages={10562--10582},
  year={2017},
  publisher={IEEE}
}

@article{bani2021behind,
  title={Behind the mask: Emotion recognition in healthcare students},
  author={Bani, Marco and Russo, Selena and Ardenghi, Stefano and Rampoldi, Giulia and Wickline, Virginia and Nowicki Jr, Stephen and Strepparava, Maria Grazia},
  journal={Medical science educator},
  volume={31},
  number={4},
  pages={1273--1277},
  year={2021},
  publisher={Springer}
}

@article{colonnello2021reduced,
  title={Reduced recognition of facial emotional expressions in global burnout and burnout depersonalization in healthcare providers},
  author={Colonnello, Valentina and Carnevali, Luca and Russo, Paolo Maria and Ottaviani, Cristina and Cremonini, Valeria and Venturi, Emanuele and Mattarozzi, Katia},
  journal={PeerJ},
  volume={9},
  pages={e10610},
  year={2021},
  publisher={PeerJ Inc.}
}

@incollection{napoli2023cost,
  title={Cost Effective Deep Learning on the Cloud},
  author={Napoli, Ot{\'a}vio O and Tesser, Rafael K and Fonseca, Daniel L and Borin, Edson},
  booktitle={High Performance Computing in Clouds: Moving HPC Applications to a Scalable and Cost-Effective Environment},
  pages={283--307},
  year={2023},
  publisher={Springer},
  address = {The Netherlands}
}

@inproceedings{kam2023devops,
  author		= "Kam, Hwee-Joo and D’Arcy, John",
  title			= "A devops perspective: The impact of role Transitions on software security continuity",
  editor		= "n/a",
  volume		= "n/a",
  booktitle		= " ECIS 2023 Proceedings ",
  pages			= "1--8",
  address		= "Kristiansand, Norway",
  publisher		= "AIS eLibrary",
  year			= "2023"
}

@article{marchand2018age,
  title={Do age and gender contribute to workers’ burnout symptoms?},
  author={Marchand, Alain and Blanc, Marie-Eve and Beauregard, Nancy},
  journal={Occupational medicine},
  volume={68},
  number={6},
  pages={405--411},
  year={2018},
  publisher={Oxford University Press UK}
}

@article{purvanova2010gender,
  title={Gender differences in burnout: A meta-analysis},
  author={Purvanova, Radostina K and Muros, John P},
  journal={Journal of vocational behavior},
  volume={77},
  number={2},
  pages={168--185},
  year={2010},
  publisher={Elsevier}
}

@article{jalili2021burnout,
  title={Burnout among healthcare professionals during COVID-19 pandemic: a cross-sectional study},
  author={Jalili, Mohammad and Niroomand, Mahtab and Hadavand, Fahimeh and Zeinali, Kataun and Fotouhi, Akbar},
  journal={International archives of occupational and environmental health},
  volume={94},
  pages={1345--1352},
  year={2021},
  publisher={Springer}
}

@article{kim2022technostress,
  title={Technostress causes cognitive overload in high-stress people: Eye tracking analysis in a virtual kiosk test},
  author={Kim, Se Young and Park, Hahyeon and Kim, Hongbum and Kim, Joon and Seo, Kyoungwon},
  journal={Information Processing \& Management},
  volume={59},
  number={6},
  pages={103093},
  year={2022},
  publisher={Elsevier}
}

@article{masri2023mental,
  title={Mental stress assessment in the workplace: a review},
  author={Masri, Ghinwa and Al-Shargie, Fares and Tariq, Usman and Almughairbi, Fadwa and Babiloni, Fabio and Al-Nashash, Hasan},
  journal={IEEE Transactions on Affective Computing},
  volume={15},
  number={3},
  pages={958--976},
  year={2023},
  publisher={IEEE},
  address = {USA}
}

@online{DevopsBurnout,
  author = {Ritvik Gupta},
  title = {{Turing.com} DevOps Burnout: Causes and Ways to Prevent It},
  year = 2024,
  url = {https://www.turing.com/blog/devops-burnout-causes-prevention},
  urldate = {2024-02-26},
  note = {last accessed, 30th of August 2024},
  organization = {Turing.com}
}

@online{greatdevopsburnout,
  author = {Anthony Molzahn},
  title = {{DevOps.com} Blogs Best of 2022: The Great DevOps Burnout},  year = 2023,
  url = {https://devops.com/the-great-devops-burnout/#disqus_thread},
  urldate = {2023-01-02},
 note = {last accessed, 30th of August 2024},
 organization  = {devops.com}
}

@online{complexitykillingdevs,
  author = {Carey, S.},
  title = {{Info World} Complexity is killing software developers},  
  year = 2021,
  url = {https://www.infoworld.com/article/2270714/complexity-is-killing-software-developers.html.},
  urldate = {2021-11-01},
  note = {last accessed, 30th of August 2024},
  publisher = {na},
  organization = {Info World}
}

@article{peticca2015perils,
  title={The perils of project-based work: Attempting resistance to extreme work practices in video game development},
  author={Peticca-Harris, Amanda and Weststar, Johanna and McKenna, Steve},
  journal={Organization},
  volume={22},
  number={4},
  pages={570--587},
  year={2015},
  publisher={Sage Publications Sage UK: London, England}
}

@inproceedings{juba2019precision,
  title={Precision-recall versus accuracy and the role of large data sets},
  author={Juba, Brendan and Le, Hai S},
  booktitle={Proceedings of the AAAI conference on artificial intelligence},
  volume={33},
  pages={4039--4048},
  year={2019},
  publisher = {na},
  address = {na}
}

@article{zhang2007comments,
  title={Comments on" data mining static code attributes to learn defect predictors"},
  author={Zhang, Hongyu and Zhang, Xiuzhen},
  journal={IEEE Transactions on Software Engineering},
  volume={33},
  number={9},
  pages={635--637},
  year={2007},
  publisher={IEEE}
}

@inproceedings{yu2018total,
  title={Total recall, language processing, and software engineering},
  author={Yu, Zhe and Menzies, Tim},
  booktitle={Proceedings of the 4th ACM SIGSOFT International Workshop on NLP for Software Engineering},
  pages={10--13},
  year={2018},
  publisher = {ACM},
  address = {USA}
}

@inproceedings{grossman2016trec,
  title={TREC 2016 Total Recall Track Overview.},
  author={Grossman, Maura R and Cormack, Gordon V and Roegiest, Adam},
  booktitle={TREC},
  year={2016},
  publisher = {na},
  address = {na},
  pages = {2369--2375}
}

@inproceedings{cormack2016scalability,
  title={Scalability of continuous active learning for reliable high-recall text classification},
  author={Cormack, Gordon V and Grossman, Maura R},
  booktitle={Proceedings of the 25th ACM international on conference on information and knowledge management},
  pages={1039--1048},
  year={2016},
  publisher = {ACM},
  address = {New York, USA}
}

@inproceedings{gray2011further,
  title={Further thoughts on precision},
  author={Gray, David and Bowes, David and Davey, Neil and Sun, Yi and Christianson, Bruce},
  booktitle={15th Annual Conference on Evaluation \& Assessment in Software Engineering (EASE 2011)},
  pages={129--133},
  year={2011},
  organization={IET},
  address = {USA},
  publisher = {IEEE}
}

@inproceedings{kasi2013cassandra,
  title={Cassandra: Proactive conflict minimization through optimized task scheduling},
  author={Kasi, Bakhtiar Khan and Sarma, Anita},
  booktitle={2013 35th International Conference on Software Engineering (ICSE)},
  pages={732--741},
  year={2013},
  organization={IEEE},
  publisher = {IEEE},
  address = {USA}
}

@inproceedings{bird2012assessing,
  title={Assessing the value of branches with what-if analysis},
  author={Bird, Christian and Zimmermann, Thomas},
  booktitle={Proceedings of the ACM SIGSOFT 20th International Symposium on the Foundations of Software Engineering},
  pages={1--11},
  year={2012},
  publisher = {ACM},
  address = {New York, USA}
}

@inproceedings{estler2014awareness,
  title={Awareness and merge conflicts in distributed software development},
  author={Estler, H Christian and Nordio, Martin and Furia, Carlo A and Meyer, Bertrand},
  booktitle={2014 IEEE 9th International Conference on Global Software Engineering},
  pages={26--35},
  year={2014},
  organization={IEEE},
  address = {USA},
  publisher = {IEEE}
}

@inproceedings{dig2007refactoring,
  title={Refactoring-aware configuration management for object-oriented programs},
  author={Dig, Danny and Manzoor, Kashif and Johnson, Ralph and Nguyen, Tien N},
  booktitle={29th International Conference on Software Engineering (ICSE'07)},
  pages={427--436},
  year={2007},
  organization={IEEE},
  publisher = {IEEE},
  address = {USA}
}

@inproceedings{bertram2010communication,
  title={Communication, collaboration, and bugs: The social nature of issue tracking in software engineering},
  author={Bertram, Dane and Voida, Amy and Greenberg, Saul and Walker, Robert},
  booktitle={Proc. ACM Conf. Comput. Support. Coop. Work},
  address = {Ney York, USA},
  publisher = {ACM},
  pages = {na},
  year={2010}
}

@article{brose2015older,
  title={Older adults’ affective experiences across 100 days are less variable and less complex than younger adults’.},
  author={Brose, Annette and De Roover, Kim and Ceulemans, Eva and Kuppens, Peter},
  journal={Psychology and aging},
  volume={30},
  number={1},
  pages={194},
  year={2015},
  publisher={American Psychological Association}
}

@article{koffer2016stressor,
  title={Stressor diversity: Introduction and empirical integration into the daily stress model.},
  author={Koffer, Rachel E and Ram, Nilam and Conroy, David E and Pincus, Aaron L and Almeida, David M},
  journal={Psychology and aging},
  volume={31},
  number={4},
  pages={301},
  year={2016},
  publisher={American Psychological Association}
}

@article{chaplin2015gender,
  title={Gender and emotion expression: A developmental contextual perspective},
  author={Chaplin, Tara M},
  journal={Emotion Review},
  volume={7},
  number={1},
  pages={14--21},
  year={2015},
  publisher={Sage Publications Sage UK: London, England}
}

@article{deng2016gender,
  title={Gender differences in emotional response: Inconsistency between experience and expressivity},
  author={Deng, Yaling and Chang, Lei and Yang, Meng and Huo, Meng and Zhou, Renlai},
  journal={PloS one},
  volume={11},
  number={6},
  pages={e0158666},
  year={2016},
  publisher={Public Library of Science San Francisco, CA USA}
}

@article{P4_destefanis2016software,
  title={Software development: do good manners matter?},
  author={Destefanis, Giuseppe and Ortu, Marco and Counsell, Steve and Swift, Stephen and Marchesi, Michele and Tonelli, Roberto},
  journal={PeerJ Computer Science},
  volume={2},
  pages={e73},
  year={2016},
  publisher={PeerJ Inc.},
  doi = {10.7717/peerj-cs.73}
}

@article{P18Ostrovsky2012,
author = { Anat   Ostrovsky  and  Joseph   Ribak  and  Avihu   Pereg  and  Dan   Gaton },
title = {Effects of job-related stress and burnout on asthenopia among high-tech workers},
journal = {Ergonomics},
volume = {55},
number = {8},
pages = {854-862},
year  = {2012},
publisher = {Taylor \& Francis},
doi = {10.1080/00140139.2012.681808},
    note ={PMID: 22676548}
}

@article{P11van2018under,
  title={Under pressure: The effects of iteration lengths on agile software development performance},
  author={van Oorschot, Kim E and Sengupta, Kishore and Van Wassenhove, Luk N},
  journal={Project Mgmt Journal},
  volume={49},
  number={6},
  pages={78--102},
  year={2018},
  publisher={SAGE Publications Sage CA: Los Angeles, CA},
  doi = {10.1177/8756972818802714}
}

@article{Haug2022,
  title={Burnout and Depression Detection Using Affective Word List Ratings},
  author={Haug S, Kurpicz-Briki M},
  journal={Studies in health technology and informatics},
  volume={292},
  number={},
  pages={43-48},
  year={2022},
  publisher={Taylor \& Francis},
  doi = {10.3233/SHTI220318}
}

@inproceedings{jacobsen1999comparison,
  title={A comparison between neural networks and decision trees},
  author={Jacobsen, Carsten and Zscherpel, Uwe and Perner, Petra},
  booktitle={International Workshop on Machine Learning and Data Mining in Pattern Recognition},
  pages={144--158},
  year={1999},
  organization={Springer},
  publisher = {Springer},
  address = {The Netherlands}
}

@article{montgomery2024comparative,
  title={A Comparative Analysis of Decision Trees, Neural Networks, and Bayesian Networks: Methodological Insights and Practical Applications in Machine Learning},
  journal  ={na},
  author={Montgomery, Richard Murdoch},
  year={2024},
  pages = {na},
  numpages = {na},
  publisher = {na},
  number = {na},
  volume = {na}
  
}

@article{verma2025explanation,
  title={Explanation of Machine Learning Algorithms Used in Disease Detection, Such as Decision Trees and Neural Networks},
  author={Verma, Nikhil and Sharma, Tripti and Kaur, Bobbinpreet},
  journal={AI in Disease Detection: Advancements and Applications},
  pages = {27--52},
  year = {2025},
  publisher = {Wiley Online Library},
  number = {na},
  volume = {na}
}

@article{bruce2009recognizing,
  title={Recognizing stress and avoiding burnout},
  author={Bruce, Susan P},
  journal={Currents in pharmacy Teaching and Learning},
  volume={1},
  number={1},
  pages={57--64},
  year={2009},
  publisher={Elsevier}
}

@inproceedings{o2008burnout,
  title={Burnout confirmed as a viable explanation for beginning teacher attrition},
  author={O'Brien, Patrick and Goddard, Richard and Keeffe, Mary},
  booktitle={Proceedings of Australian association for research in education annual conference (AARE 2007)},
  year={2008},
  publisher = {na},
  address = {na},
  pages = {na},
  numpages = {na}
}

@article{chakrabarti2022more,
  title={More than burnout: qualitative study on understanding attrition among senior Obstetrics and Gynaecology UK-based trainees},
  author={Chakrabarti, Rima and Markless, Sharon},
  journal={Bmj Open},
  volume={12},
  number={2},
  pages={e055280},
  year={2022},
  publisher={British Medical Journal Publishing Group}
}

@article{stehman2019burnout,
  title={Burnout, drop out, suicide: physician loss in emergency medicine, part I},
  author={Stehman, Christine R and Testo, Zachary and Gershaw, Rachel S and Kellogg, Adam R},
  journal={Western Journal of Emergency Medicine},
  volume={20},
  number={3},
  pages={485},
  year={2019}
}

@article{ahsan2021effect,
  title={Effect of data scaling methods on machine learning algorithms and model performance},
  author={Ahsan, Md Manjurul and Mahmud, MA Parvez and Saha, Pritom Kumar and Gupta, Kishor Datta and Siddique, Zahed},
  journal={Technologies},
  volume={9},
  number={3},
  pages={52},
  year={2021},
  publisher={MDPI}
}

@article{de2023choice,
  title={The choice of scaling technique matters for classification performance},
  author={de Amorim, Lucas BV and Cavalcanti, George DC and Cruz, Rafael MO},
  journal={Applied Soft Computing},
  volume={133},
  pages={109924},
  year={2023},
  publisher={Elsevier}
}

@inproceedings{reiss2012introducing,
  title={Introducing a new benchmarked dataset for activity monitoring},
  author={Reiss, Attila and Stricker, Didier},
  booktitle={2012 16th international symposium on wearable computers},
  pages={108--109},
  year={2012},
  organization={IEEE},
  publisher = {IEEE},
  address = {USA}
}

@article{jimenez2023swe,
  title={Swe-bench: Can language models resolve real-world github issues?},
  author={Jimenez, Carlos E and Yang, John and Wettig, Alexander and Yao, Shunyu and Pei, Kexin and Press, Ofir and Narasimhan, Karthik},
  journal={arXiv preprint arXiv:2310.06770},
  year={2023},
  publisher = {na},
  address = {na},
  number = {na},
  volume = {na},
  pages = {na},
  numpages = {na}
}

@inproceedings{doan2023too,
  title={Too long; didn’t read: Automatic summarization of github readme. md with transformers},
  author={Doan, Thu TH and Nguyen, Phuong T and Di Rocco, Juri and Di Ruscio, Davide},
  booktitle={Proceedings of the 27th International Conference on Evaluation and Assessment in Software Engineering},
  pages={267--272},
  year={2023},
  publisher = {ACM},
  address = {USA}
}

@article{ali2012random,
  title={Random forests and decision trees},
  author={Ali, Jehad and Khan, Rehanullah and Ahmad, Nasir and Maqsood, Imran},
  journal={International Journal of Computer Science Issues (IJCSI)},
  volume={9},
  number={5},
  pages={272},
  year={2012}
}

@article{rey2016emotional,
  title={Emotional competence relating to perceived stress and burnout in Spanish teachers: a mediator model},
  author={Rey, Lourdes and Extremera, Natalio and Pena, Mario},
  journal={PeerJ},
  volume={4},
  pages={e2087},
  year={2016},
  publisher={PeerJ Inc.}
}

@article{spiller2021emotion,
  title={Emotion network density in burnout},
  author={Spiller, Tobias R and Weilenmann, Sonja and Prakash, Krithika and Schnyder, Ulrich and von K{\"a}nel, Roland and Pfaltz, Monique C},
  journal={BMC psychology},
  volume={9},
  pages={1--9},
  year={2021},
  publisher={Springer}
}

@article{zapf2002emotion,
  title={Emotion work and psychological well-being: A review of the literature and some conceptual considerations},
  author={Zapf, Dieter},
  journal={Human resource management review},
  volume={12},
  number={2},
  pages={237--268},
  year={2002},
  publisher={Elsevier}
}

@article{kleiner2017oncologist,
  title={Oncologist burnout and compassion fatigue: investigating time pressure at work as a predictor and the mediating role of work-family conflict},
  author={Kleiner, Sibyl and Wallace, Jean E},
  journal={BMC health services research},
  volume={17},
  pages={1--8},
  year={2017},
  publisher={Springer}
}

@article{dreison2018integrating,
  title={Integrating self-determination and job demands--resources theory in predicting mental health provider burnout},
  author={Dreison, Kimberly C and White, Dominique A and Bauer, Sarah M and Salyers, Michelle P and McGuire, Alan B},
  journal={Administration and Policy in Mental Health and Mental Health Services Research},
  volume={45},
  pages={121--130},
  year={2018},
  publisher={Springer}
}

@article{makara2019self,
  title={Self-efficacy as a moderator between stress and professional burnout in firefighters},
  author={Makara-Studzi{\'n}ska, Marta and Golonka, Krystyna and Izydorczyk, Bernadetta},
  journal={International journal of environmental research and public health},
  volume={16},
  number={2},
  pages={183},
  year={2019},
  publisher={MDPI}
}

@article{ventura2015professional,
  title={Professional self-efficacy as a predictor of burnout and engagement: The role of challenge and hindrance demands},
  author={Ventura, Mercedes and Salanova, Marisa and Llorens, Susana},
  journal={The Journal of psychology},
  volume={149},
  number={3},
  pages={277--302},
  year={2015},
  publisher={Taylor \& Francis}
}
%

\end{document}